\documentclass[12pt,preprint]{aastex}
\usepackage{graphicx}

\def\be{\begin{equation}}
\def\ee{\end{equation}}
\def\ba{\begin{eqnarray}}
\def\ea{\end{eqnarray}}
\newcommand{\msun}{\ifmmode\mbox{M}_{\odot}\else$\mbox{M}_{\odot}$\fi}
\newcommand{\rsun}{\ifmmode\mbox{R}_{\odot}\else$\mbox{M}_{\odot}$\fi}
\newcommand{\degrees}{\ifmmode^{\circ}\else$^{\circ}$\fi}
\newcommand{\degree}{\ifmmode^{\circ}\else$^{\circ}$\fi}
\newcommand{\amin}{\ifmmode^{\prime}\else$^{\prime}$\fi}
\newcommand{\asec}{\ifmmode^{\prime\prime}\else$^{\prime\prime}$\fi}

\shorttitle{New Discoveries from AO327}
\shortauthors{Deneva et al.}

\begin{document}

\title{New Discoveries from the Arecibo 327~MHz Drift Pulsar Survey Radio Transient Search} 
\author{
J.~S.~Deneva$^{1,*}$, 
K.~Stovall$^{2}$,
M.~A.~McLaughlin$^{3}$,
M.~Bagchi$^{3,4}$,
S.~D.~Bates$^{4}$,
P.~C.~C.~Freire$^{5}$,
J.~G.~Martinez$^{5,6}$,
F.~Jenet$^{6}$,
N.~Garver-Daniels$^{3}$}

\affil{$^{1}$National Research Council, resident at the Naval Research Laboratory, Washington, DC 20375}
\affil{$^{2}$Department of Physics and Astronomy, University of New Mexico, Albuquerque, NM 87131}
\affil{$^{3}$Department of Physics and Astronomy, West Virginia University, Morgantown, WV 26506}
\affil{$^{4}$The Institute of Mathematical Sciences, Chennai, India 600113}
\affil{$^{5}$Max-Planck-Institut f\"{u}r Radioastronomie, Bonn, Germany}
\affil{$^{6}$Center for Advanced Radio Astronomy, Department of Physics and Astronomy, University of Texas at Brownsville, Brownsville, TX 78520}

\begin{abstract}
We present Clusterrank, a new algorithm for identifying dispersed astrophysical pulses. Such pulses are commonly detected from Galactic pulsars and rotating radio transients (RRATs), which are neutron stars with sporadic radio emission. { More recently, isolated, highly dispersed pulses dubbed fast radio bursts (FRBs) have been identified as the potential signature of an extragalactic cataclysmic radio source distinct from pulsars and RRATs.} Clusterrank helped us discover 14 pulsars and 8 RRATs in data from the Arecibo 327~MHz Drift Pulsar Survey (AO327). The new RRATs have DMs in the range $23.5 - 86.6$~pc~cm$^{-3}$ and periods in the range $0.172 - 3.901$~s. The new pulsars have DMs in the range $23.6 - 133.3$~pc~cm$^{-3}$ and periods in the range $1.249 - 5.012$~s, and include two nullers and a mode-switching object. We estimate an upper limit on the all-sky FRB rate of $10^5$~day$^{-1}$ for bursts with a width of 10~ms and flux density $\gtrsim 83$~mJy. The DMs of all new discoveries are consistent with a Galactic origin. In comparing statistics of the new RRATs with sources from the RRATalog, we find that both sets are drawn from the same period distribution. In contrast, we find that the period distribution of the new pulsars is different from the period distributions of canonical pulsars in the ATNF catalog or pulsars found in AO327 data by a periodicity search. This indicates that Clusterrank is a powerful complement to periodicity searches and uncovers a subset of the pulsar population that has so far been underrepresented in survey results and therefore in Galactic pulsar population models. 
\end{abstract}

\section{Introduction}

The field of fast radio transient detection as a means of discovering new radio sources first came to the forefront when \cite{McLaughlin06} found 11 such transients in archival Parkes Multibeam Survey data. They were called Rotating Radio Transients (RRATs) as the differences between pulse arrival times for each object were found to be multiples of one interval, the rotation period. RRAT rotation periods are on the order of a few hundreds to a few thousands of milliseconds, consistent with rotating neutron stars. The average RRAT rotation period is larger than the average normal pulsar rotation period. However, for some RRATs detected in only one or two observations the published period may be multiples of the actual period because of the small number of pulses detected. Furthermore, there are observational selection effects which result in pulsars with longer periods being detected with higher signal-to-noise in single-pulse searches \citep{McLaughlin03}.

Unlike normal pulsars, RRATs appear to not be emitting a pulse on every rotation, as evidenced by the fact that these objects were missed by Fast Fourier Transform-based periodicity searches. As more RRATs were discovered (see the RRATalog\footnote{\tt http://astro.phys.wvu.edu/rratalog}) and more follow-up observations accumulated, the diversity in emission patterns has made it increasingly likely that different processes are responsible for the intermittency of what initially appeared as one new class of radio sources. 

Some objects discovered by single-pulse searches are slow pulsars selected against in surveys with short integration times where there are not enough pulses for a detection to be made via periodicity search. Some RRATs discovered e.g. at 1.4~GHz appear as normal pulsars when observed at a lower frequency \citep{Deneva09}. This is consistent with the explanation of \cite{Weltevrede06} that in such cases the intermittency is due to a pulse intensity distribution with a long high-flux tail, such that as the pulsar flux density drops off at higher frequencies, only the brightest pulses remain detectable. RRATs emitting short sequences of pulses on consecutive rotations may be extreme nullers and/or old pulsars close to the death line, where the radio emission mechanism gradually begins to turn off (\citealt{Zhang07}, \citealt{Burke10}). In other cases, a single pulse or a single short sequence of pulses is detected and the RRAT is never seen again despite many follow-up observations \citep{Deneva09}. These detections are consistent with cataclysmic events or a mechanism which generates rare conditions in an otherwise quiescent neutron star magnetosphere. \cite{Cordes08} argue that this emission pattern can be explained by sporadic accretion of debris from a circumpulsar asteroid belt. 

All RRATs known to date have dispersion measures (DMs, the integrated column density of ionized gas along the line of sight) consistent with a Galactic origin. \cite{Lorimer07} reported a 1.4~GHz Parkes detection of a fast radio transient outside of the Galactic plane with a DM significantly exceeding the estimated contribution of Galactic ionized gas along the line of sight. More Parkes detections of transients with similar properties were made by \cite{Thornton13}, \cite{Petroff15}, and \cite{Ravi15}, the latter from a targeted observation of the Carina dwarf spheroidal galaxy. \cite{Spitler14} detected a transient with similar properties using Arecibo, also at 1.4~GHz. {Apart from their high DMs, most of these fast radio bursts (FRBs) differ from RRATs in that they are only detected with one pulse in their discovery observations. Despite many follow-up observations, so far repeat pulses have been definitively detected only from the Arecibo FRB \citep{Spitler16}.} The combination of seemingly extragalactic origin and, until recently, the lack of repeat bursts has suggested cataclysmic events producing a single coherent radio pulse detectable to Gpc distances, such as coalescing neutron stars \citep{Hansen01}, evaporating black holes \citep{Rees77}, or collapsing supramassive neutron stars \citep{Falcke14}. Dissenting views have attributed FRBs to Galactic flaring stars (\citealt{Loeb14}, \citealt{Maoz15}) and atmospheric phenomena \citep{Kulkarni14}. 

Understanding the nature of fast radio transients is important in figuring out how their progenitors fit into the evolution of our Galaxy and galaxies in general. They may even provide an independent test of various evolutionary scenarios. For example, if RRATs are assumed to be intermittent since formation and comprise a neutron star population separate from normal pulsars, the Galactic core-collapse supernova rate is too low to account for both populations \cite{kk08}. This is not the case if RRATs represent a stage in the evolution of pulsars, even though they may outnumber other pulsar types since their sporadic pulses make them less likely to be discovered by pulsar surveys. All FRBs known to date have been found at 1.4~GHz even though surveys conducted at 350 MHz with the Green Bank telescope search DMs up to 1000~pc~cm$^{-3}$ \citep{Karako15}. It is still unknown whether that is due only to selection effects or has intrinsic causes as well. 

In this paper we report the results of running Clusterrank, a new algorithm for identifying astrophysical radio transients, on data collected by the Arecibo 327~MHz Drift Pulsar Survey (AO327). Section~\ref{sec_obs} describes the AO327 survey setup and observations, Section~\ref{sec_sp} gives details on the single-pulse search code whose output Clusterrank operates on, and Section~\ref{sec_clusterrank} focuses on Clusterrank implementation and performance. Sections~\ref{sec_psrs} and \ref{sec_rrats} present new pulsars and RRATs, respectively, and Section~\ref{sec_stats} analyzes the statistics of both types of discoveries. Finally, Section~\ref{sec_frblimits} places limits on the FRB population. 

\section{AO327 Survey Observations}\label{sec_obs}

The AO327 drift survey is running since 2010 during Arecibo telescope downtime or unassigned time. It aims to search the entire Arecibo sky (declinations from $-1\degree$ to $38\degree$) for pulsars and transients at 327~MHz. Phase I of the survey covers declinations from $-1\degree$ to $28\degree$, and Phase II will cover the remainder of the sky accessible to Arecibo. Under normal operating conditions, AO327 does not get observing time within $\pm 5\degree$ from the Galactic plane. Frequencies higher than 327~MHz are more suitable for pulsar and transient searches within the Galactic plane because of significant dispersion and scattering due to Galactic ionized gas. However, telescope time occasionally becomes available on short notice due to technical problems that render regularly scheduled projects unable to observe. AO327 is a filler project in such cases, and some of its discoveries were made during unscheduled encroachments on the Galactic plane. 

{In this paper we present single-pulse search results from analyzing 882~h of data taken with the Arecibo 327~MHz receiver and the Mock spectrometer backend, up until March 2014, when AO327 began using the newer PUPPI backend. An analysis of PUPPI data will be presented in a future paper.}
%Including Mock and PUPPI observations, at the time of writing the Phase I survey area (declinations from -1\degree\ to 28\degree) is XXX\% complete. 
The effective integration time is $T_{\rm obs} = 60$~s for AO327 observations, corresponding to the drift time through the beam at 327~MHz. For Mock observations, the number of channels is $N_{\rm ch} = 1024$, the bandwidth is $\Delta\nu = 57$~MHz, the sampling time is $dt = 125~\mu$s, the receiver temperature is $T_{\rm rec} = 115$~K, and the gain is $G = 11$~K/Jy. 

\begin{figure}[t]
\begin{center}
\includegraphics[width=\textwidth]{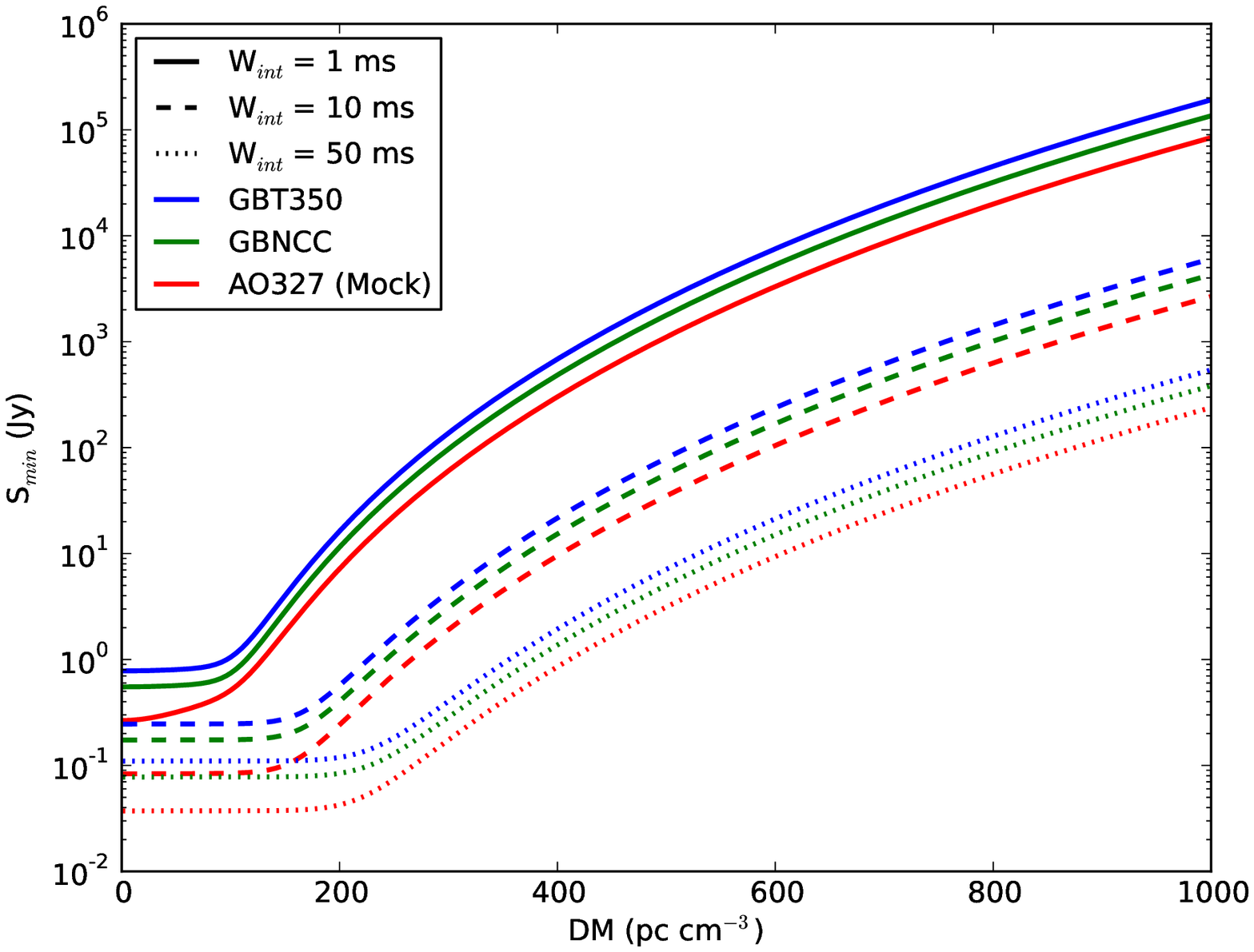}
\caption{The minimum detectable single-pulse flux density $S_{\rm min}$ vs. DM for AO327 using the Mock spectrometer (red), GBNCC (green), and GBT350 (blue), for several intrinsic pulse widths $W_{\rm int}$. The inflection point in each curve corresponds to the transition from dispersion-limited to scattering-limited detection regime. \label{fig_smin}}
\end{center}
\end{figure}

Figure~\ref{fig_smin} shows the intrinsic minimum detectable flux density $S_{\rm int,min}$ vs. DM for AO327 using the Mock spectrometer, for the Green Bank North Celestial Cap survey (GBNCC, \citealt{Stovall14}), and for the GBT350 drift survey \citep{Lynch13}. According to the radiometer equation {applied to single pulse detection \citep{Cordes03}}
\be
S_{\rm int,min} = \left(\frac{W_{\rm obs}}{W_{\rm int}}\right)~\frac{\left(T_{\rm rec} + T_{\rm sky}\right)~SNR_{\rm min}}{G \left(N_{\rm pol}~\Delta\nu~W_{\rm obs}\right)^{1/2}},\label{eqn_smin}
\ee
where $SNR_{\rm min} = 6$ is the detection threshold, the sky temperature $T_{\rm sky} = 50$~K \citep{Haslam82} (appropriate for a source out of the Galactic plane), $W_{\rm int}$ is the intrinsic pulse width, and $W_{\rm obs}$ is the observed broadened pulse width. The two pulse width quantities are related by 
\be
W_{\rm obs} = \left(W_{\rm int}^2 + dt^2 + \tau_s^2 + \Delta t_{\rm DM,1ch}^2\right)^{1/2},
\ee
where $\tau_s$ is the scattering broadening estimated from Eqn.~7 in \cite{Bhat04}, and $\Delta t_{\rm DM,1ch}$ is the dispersion delay across a channel width. \cite{Deneva13} present a more detailed discussion of AO327 search volume, sensitivity to periodic sources, and comparisons with other pulsar surveys as well as between the different backends that have been used in AO327 observations. 

%System parameters of AO327-Mock (for calculating flux densities from discoveries):
%Nch = 1024
%BW = 57.34 MHz
%dt = 0.000125 s
%Trec = 115 K 
%Gain = 11 K/Jy
%Tsky ~ 50 K (from Haslam 1982)

\section{Single-Pulse Search}\label{sec_sp}

Data are dedispersed with 6358 trial DMs in the range $0 - 1095$~pc~cm$^{-3}$. The spacing between successive trial DMs increases from 0.02 to 1.0~pc~cm$^{-3}$ such that at high DMs the smearing due to a pulsar's actual DM being halfway between two trial DMs is much smaller than the scattering broadening estimated from the empirical fit of \cite{Bhat04}. Because scattering broadening dwarfs the sampling time even at moderate DMs, during dedispersion data are downsampled by a factor that increases as the trial DM increases (Table~\ref{tab_dedisp}). 

\begin{table}[t]
\begin{center}
\caption{The step between successive trial DMs and the downsampling factors used for different DM ranges in processing AO327 Mock data. As the trial DM increases, uncorrectable scattering broadening begins to dominate sensitivity. The progressively increasing DM spacing and downsampling factor are chosen such that computational efficiency is maximized for each DM range while the increased dispersion smearing is still negligible compared to scattering broadening. \label{tab_dedisp}}
\begin{tabular}{rrrr}
\hline
Low DM & High DM & DM step & $N_{\rm downsamp}$ \\
(pc~cm$^{-3}$) & (pc~cm$^{-3}$) & (pc~cm$^{-3}$) & \\
\hline
0.00   & 36.94  & 0.02    & 1 \\
36.96  & 58.35   & 0.03   & 2 \\
58.38  & 99.13   & 0.05   & 4 \\
99.18  & 201.08  & 0.10   & 8 \\
201.18 & 482.88  & 0.30   & 16 \\
483.18 & 890.68  & 0.50   & 32 \\
891.18 & 1095.18 & 1.00   & 64 \\
\hline
\end{tabular}
\end{center}
\end{table}

{We use the PRESTO\footnote{\tt http://www.cv.nrao.edu/\~{}sransom/presto} tool {\tt single\_pulse\_search.py}} to search each dedispersed time series for pulses. Each radio pulse, astrophysical or terrestrial, is typically detected as a cluster of events above a signal-to-noise threshold at multiple closely spaced trial DMs. We use the word ``event'' to refer to such a detection at a single trial DM. The one-dimensional time series are flattened with a piecewise linear fit where each piece is 1000 bins long. Then the time series are convolved with a set of boxcar functions with widths ranging from 1 to 300 bins. Because the time series may already have been downsampled during dedispersion, the same boxcar function may correspond to different absolute widths in seconds for different time series. We cap the width of boxcars such that they do not exceed 0.1~s for any time series. This corresponds to the maximum pulse duration detectable in our search. Observed RRAT and FRB pulse widths range from a fraction of a millisecond to a few tens of milliseconds\footnote{\tt http://astro.phys.wvu.edu/rratalog}$^{,}$\footnote{\tt http://astro.phys.wvu.edu/FRBs/FRBs.txt}. We construct a list of events with a signal-to-noise ratio ($SNR$) $\geq 5$ for each time series, where 
\be
SNR = \frac{{\displaystyle \sum_{i}}{\left(S_i - S_0\right)}}{\sigma W_{\rm box}^{1/2}}.
\ee
The sum is over successive bins $S_i$ covered by the boxcar function, $S_0 \approx 0$ is the baseline level after flattening, $\sigma \approx 1$ is the root-mean-square noise after normalization, and $W_{\rm box}$ is the boxcar width in number of bins. This definition of $SNR$ has the advantage that it gives approximately the same result regardless of the downsampling factor used for the time series, as long as the pulse is still resolved. If there are several events with $SNR > 5$ detected with boxcars of different widths from the same portion of data, only the event with the highest $SNR$ is retained in the final list. 

The scattering broadening observed for most known FRBs (DMs~$\sim 500 - 1000$~pc~cm$^{-3}$) is $\lesssim 1$~ms at 1400~MHz. Assuming a Kolmogorov scattering spectrum such that $\tau_s \propto f^{-4}$, this corresponds to a scattering time of $\sim 300$~ms at 327~MHz. The widest boxcar template PRESTO uses for event detection in dedispersed time series is 150 bins. At the maximum downsampling factor of 64, with our sampling time of $81.92~\mu$s, this corresponds to a template width of $\sim 800$~ms. Assuming an intrinsic pulse width of 5~ms, a 6-$\sigma$ detection of a pulse with $W_{\rm obs} = 800$~ms corresponds to $S_{\rm int} = 1.5$~Jy, and a similar detection of a pulse with $W_{\rm obs} = 300$~ms corresponds to $S_{\rm int} = 0.9$~Jy. 

\begin{figure}[t]
\begin{center}
\includegraphics[width=\textwidth]{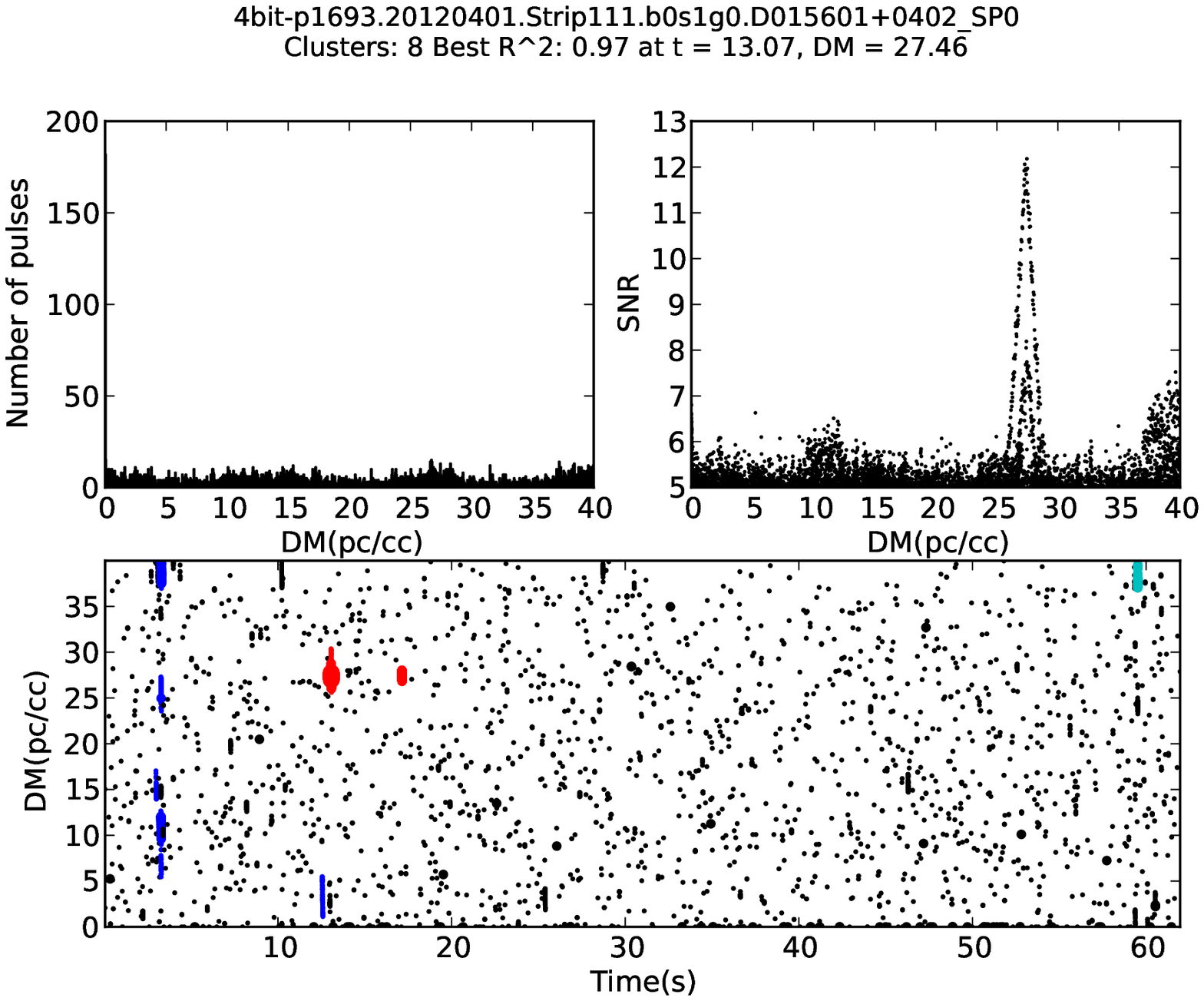}
\caption{Single-pulse search plot of the discovery observation of RRAT J0156+04. Top: Histograms of the number of events vs. DM (left) and event $SNR$ vs. DM (right). Bottom: Events are plotted vs. DM and time. Larger marker sizes correspond to higher $SNR$. Events belonging to clusters identified by Clusterrank are shown in red if the cluster $R^2 > 0.8$, {magenta if $0.7 < R^2 \leq 0.8$, cyan if $0.6 < R^2 \leq 0.7$, green if $0.5 < R^2 \leq 0.6$, and blue if $R^2 \leq 0.5$. (There are no clusters with $0.7 < R^2 \leq 0.8$ or $0.5 < R^2 \leq 0.6$ in this case.)} The two clusters of events shown in red correspond to the two superimposed peaks in the $SNR$ vs. DM histogram on upper right. The plot title identifies that the cluster whose $SNR$ vs. DM signature most closely matches Eqn.~\ref{eqn_lsqfit} has $R^2 = 0.97$, the arrival time of the highest-$SNR$ event within the cluster is $t = 13.07$~s since the start of the data span shown, and the best-fit DM is 27.46~pc~cm$^{-3}$. \label{fig_0156spplot}}
\end{center}
\end{figure}

The event list produced by PRESTO is used to make plots like Figure~\ref{fig_0156spplot}, which are then inspected by eye {to look for clusters of events indicative of dispersed pulses. Because the spacing between trial DMs changes significantly within the full range of DMs used in the search (Table~\ref{tab_dedisp}), single-pulse search plots are made for four subsets of the full range of trial DMs.} AO327 data are processed in 1-minute ``beams'', corresponding to the maximum transit time through the Arecibo beam at 327~MHz. To date, we have processed $\sim 882$~h of Mock drift data, resulting in a total of 423360 single-pulse search plots. 
%{\bf (XXX: Kevin and Maura reported they have processed 1100 and 168 h, respectively. I have SP files for 882 h of Mock data, and thought that was from the entire processed Mock data set. Were SP files not made for some beams?)} 
Because AO327 is a blind all-sky survey, the vast majority of these plots contain only events due to Gaussian noise or radio frequency interference (RFI). Since human inspection of all single-pulse search plots would require an excessive amount of time, ideally we want this task to be reliably accomplished by an algorithm able to distinguish astrophysical dispersed pulses from terrestrial RFI or noise, in a constantly changing RFI environment. Below we describe such an algorithm, called Clusterrank, that enabled us to quickly discover 22 new pulsars and RRATs.

\section{Clusterrank}\label{sec_clusterrank}

Clusterrank\footnote{\tt http://github.com/juliadeneva/clusterrank} operates on the event lists produced by the PRESTO single-pulse search for a 1-minute span of AO327 drift data. Events are sorted by DM and time and clusters of events are identified such that the DM and time gaps between sorted events do not exceed a threshold. We have set the maximum acceptable DM gap to 1~pc~cm$^{-3}$, the largest spacing in our trial DM list. We use a maximum acceptable time gap corresponding to the product of the raw data time resolution and the largest boxcar function width used in the single-pulse search: 0.125~ms $\times$ 150 samples $\approx 19$~ms. The minimum number of events per cluster that would trigger further processing is set to 50. The DM gap, time gap, and minimum events per cluster are tunable parameters and the values chosen for processing the AO327 Mock data set strike a balance between detecting as many clusters likely to be caused by astrophysical pulses as possible and avoiding further processing of the excessive number of smaller clusters occurring randomly due to Gaussian noise fluctuations. Similar to PRESTO's {\tt single\_pulse\_search.py}, Clusterrank considers and plots four separate DM ranges: $0 - 40$~pc~cm$^{-3}$, $30 - 120$~pc~cm$^{-3}$, $100 - 500$~pc~cm$^{-3}$, and $500 - 1000$~pc~cm$^{-3}$.

The determination of how likely a cluster is to indicate the presence of a dispersed pulse in the data hinges on the analytical expression describing how a pulse's amplitude in the dedispersed time series changes as the trial DM varies with respect to the actual pulsar DM. \cite{Cordes03} derive the ratio of the peak flux density of a Gaussian pulse dedispersed with a DM error $\delta$DM to the peak flux density if the same pulse is dedispersed with no DM error. {We substitute event $SNR$ for the peak flux density. The DM error $\delta{\rm DM}_{\rm i} = {\rm DM}_{\rm i} - {\rm DM}_{\rm psr}$, where the index $i$ refers to an event in the cluster. If $SNR_{\rm psr}$ is the pulse SNR for $\delta{\rm DM}=0$, the resulting equations are 
\be
\frac{SNR\left(\delta {\rm DM_i}\right)}{SNR_{\rm psr}} = \frac{\sqrt{\pi}}{2}\zeta^{-1}{\rm Erf}\left(\zeta\right), \label{eqn_lsqfit}
\ee
where
\be
\zeta = 6.91 \times 10^{-3} \delta {\rm DM_i} \frac{\Delta\nu_{\rm MHz}}{W_{\rm ms}~\nu^3_{\rm GHz}}. \label{eqn_zeta}
\ee
Here $\Delta\nu_{\rm MHz}$ is the bandwidth in MHz, $W_{\rm ms}$ is the observed pulse width in ms, and $\nu_{\rm GHz}$ is the center observing frequency in GHz. We perform least-squares fitting using the the {\tt Optimize.leastsq} module of SciPy and the recorded $SNR$s and DMs of the events in a cluster. The free parameters in the fit are $W_{\rm ms}$, ${\rm DM_ {psr}}$, and $SNR_{\rm psr}$. }The initial guess values passed to the least-squares fitting function are 10~ms as the width, and the DM and $SNR$ of the event with the highest $SNR$ in the cluster. 

\begin{figure}[t]
\begin{center}
\includegraphics[width=0.7\textwidth]{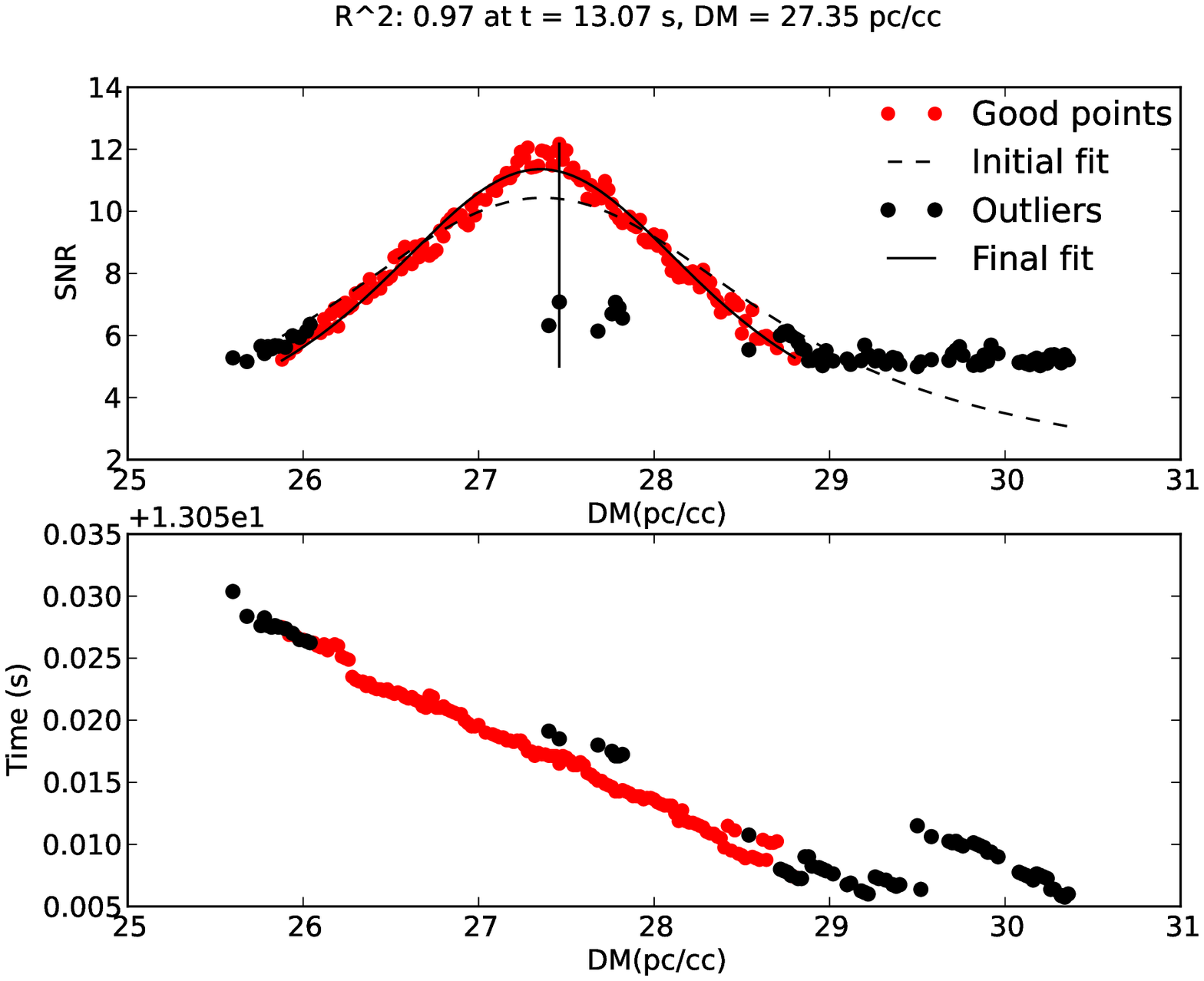}
\caption{Clusterrank fit for the brigher pulse of RRAT J0156+04 from the discovery observation shown in Figure~\ref{fig_0156spplot}. Top: a dashed curve shows the best fit of $SNR$ vs. DM without outlier rejection. A solid curve shows the best fit after three iterations of identifying outliers, removing them from the cluster, and redoing the fit. Good points used in the final fit are shown in red, and outliers are shown in black. A vertical line is drawn through the event with the highest $SNR$, whose $SNR$ and DM are used as seeds for the initial least-squares fit. There are several outliers at low $SNR$ close to the best-fit DM, indicating a two-peak pulse shape, with one component significantly weaker than the other. {Bottom: the structure of the cluster is shown in DM-time space.}
%{\bf (XXX: How to explain the tail of outliers on the right? A lot of pulses have such a tail on one or both sides.)} Bottom: the structure of the cluster is shown in DM-time space. 
\label{fig_0156pulse1}}
\end{center}
\end{figure}

\begin{figure}[t]
\begin{center}
\includegraphics[width=0.7\textwidth]{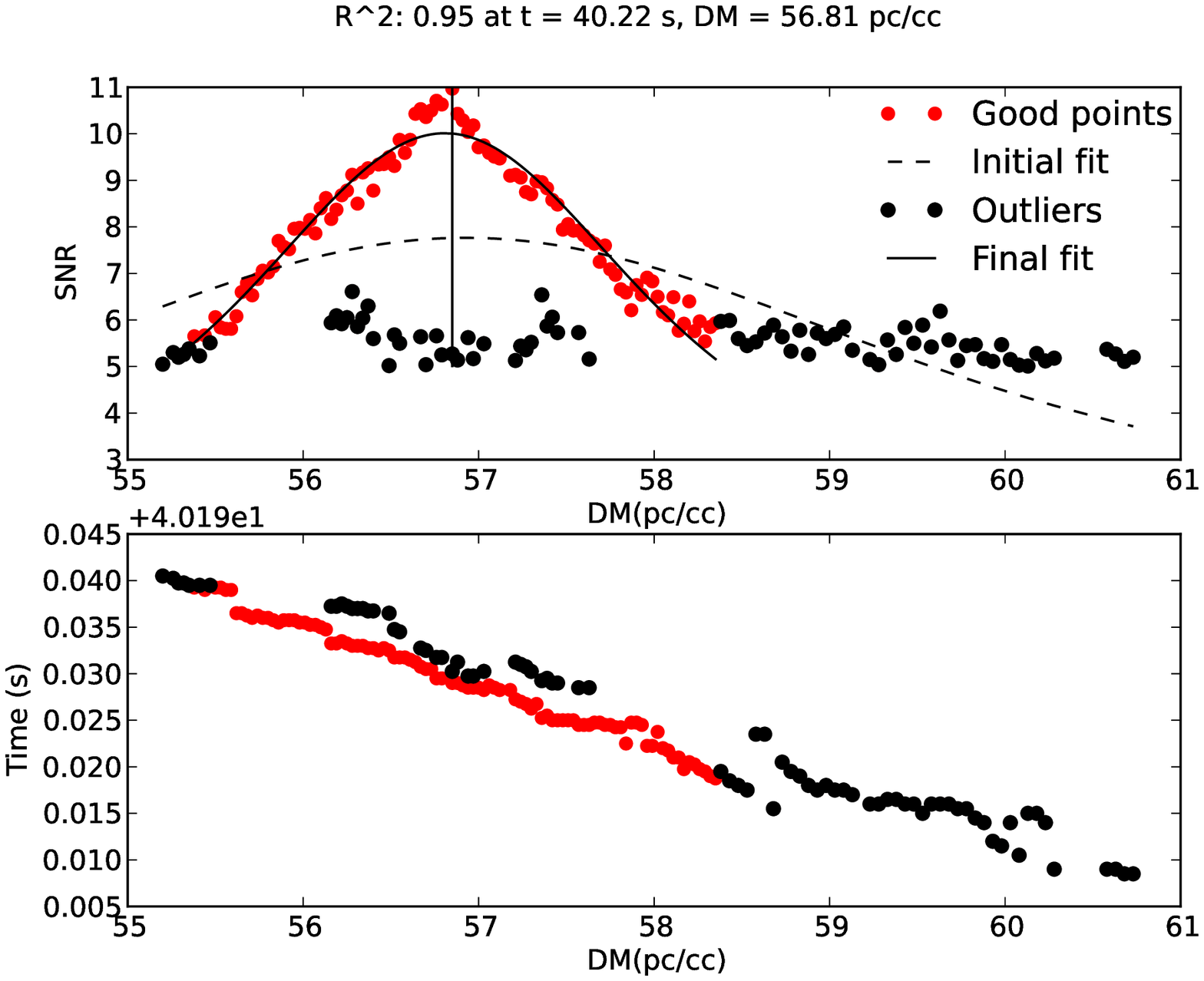}
\caption{Clusterrank fit for the single pulse in the discovery observation of RRAT J0544+20. Top: a dashed curve shows the best fit of $SNR$ vs. DM without outlier rejection. A solid curve shows the best fit after three iterations of identifying outliers, removing them from the cluster, and redoing the fit. Good points used in the final fit are shown in red, and outliers are shown in black. A vertical line is drawn through the event with the highest $SNR$, whose $SNR$ and DM are used as seeds for the initial least-squares fit. There are several outliers at low $SNR$ close to the best-fit DM, indicating a two-peak pulse shape, with one component significantly weaker than the other. Bottom: the structure of the cluster is shown in DM-time space. \label{fig_0544pulse1}}
\end{center}
\end{figure}

Due to pulse substructure, noise, as well as the imperfect selection of a best-width boxcar filter for pulses with low $SNR$, in a cluster of events there are often outliers that significantly deviate from the $SNR$ vs. trial DM dependence predicted by Eqn.~\ref{eqn_lsqfit} and Eqn.~\ref{eqn_zeta}. We perform three iterations of identifying outliers, removing them from the cluster, and redoing the least-squares fit with the remaining events. {An event is rejected as an outlier if $|SNR(\delta {\rm DM_i}) - SNR_{\rm i}| > |SNR(\delta {\rm DM_i}) - 5|/2$. The baseline SNR difference of 5 was chosen to correspond to the minimum SNR for which events are recorded. In absolute terms, the rejection criterion is more stringent for events further away in DM from the peak in SNR vs. DM space. This effectively rejects the flat tails at SNR = 5 exhibited by many clusters.} Figures \ref{fig_0156pulse1} and \ref{fig_0544pulse1} show the resulting improvement in the final fit for two clusters of events containing outliers and the effect outliers can have on the quality of the initial fit. {The bottom panels of the two figures show the sloping signature of the event clusters in time-DM space. This is due to dispersion under- or overcorrection away from the actual RRAT DM smearing the pulse and shifting its peak in the dedispersed time series to a later or earlier time, respectively.}

\subsection{Test Statistic}\label{sec_ts}

As a measure of the goodness of fit for each cluster we use the correlation coefficient for the final fit of the cluster events $SNR$ vs. DM
\be
R^2 = 1 - \frac{\sum \left(SNR_i - \widehat{SNR}_i \right)^2}{\sum \left(SNR_i - \overline{SNR} \right)^2},
\ee
where $SNR_i$ is the $i$-th event's $SNR$, $\widehat{SNR}_i$ is the $i$-th event's predicted $SNR$ based on Eqn.~\ref{eqn_lsqfit}, and $\overline{SNR}$ is the mean $SNR$ for the cluster. The number of events can vary widely from one cluster to another and we find that in this situation $R^2$ is a better test statistic than the reduced $\chi^2$ or the root-mean-square residual from the least-squares fit. Hereafter we use the term ``score'' to refer to the $R^2$ value of a cluster. 

After all clusters in one of the considered DM ranges are fitted, the highest score for that DM range and beam is recorded. Plots are viewed in decreasing order of the recorded best score values. The range of possible values for $R^2$ is from zero (no correlation between cluster events and fit) to unity (perfect correlation). We find that pulses from known pulsars that would be unambiguously identified as such on visual inspection when viewed in isolation from other pulses of the same pulsar are almost always fitted with $R^2 > 0.9$, and the remainder are fitted with $0.8 < R^2 < 0.9$. We therefore adopt $R^2 > 0.8$ as the threshold for visual inspection of plots. 

\begin{figure}[t]
\begin{center}
\includegraphics[width=\textwidth]{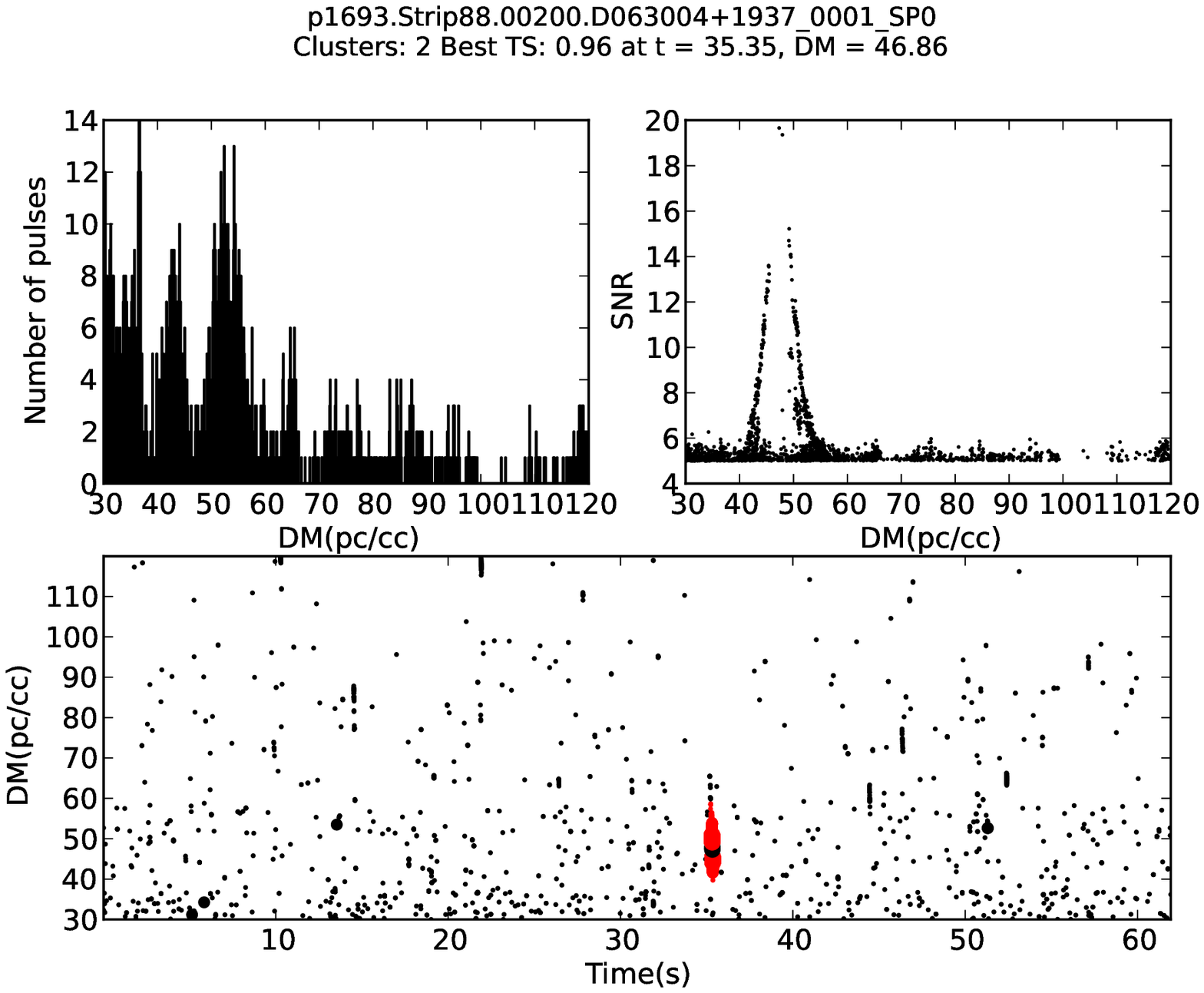}
\caption{Single-pulse search plot of the discovery observation of PSR~J0630+19. Histograms of the number of events vs. DM (left) and event $SNR$ vs. DM (right). Bottom: Events are plotted vs. DM and time. Larger marker sizes correspond to higher $SNR$. Events belonging to clusters identified by Clusterrank are shown in red if the cluster $R^2 > 0.8$. {In this case, the pulse yields three clusters of events at $t \sim 35$~s, with the cluster corresponding to the peak in $SNR$ vs. DM space consisting of only two events and therefore not fitted.} The discovery of this pulsar was made based on the fits of the two shoulders of the $SNR$ vs. DM signature of the pulse, detected as two separate clusters with scores of 0.96 and 0.81 (Figure~\ref{fig_0630pulse1} and Figure~\ref{fig_0630pulse2}). {The clusters at $DM \sim 53$~pc~cm$^{-3}$, $t \sim 13$ and 52~s were not fitted because they contain too few events. They are unlikely to be pulses from PSR~J0630+19 since their DM deviates significantly from the pulsar DM of 48~pc~cm$^{-3}$.} \label{fig_0630spplot}}
\end{center}
\end{figure}

\begin{figure}[t]
\begin{center}
\includegraphics[width=\textwidth]{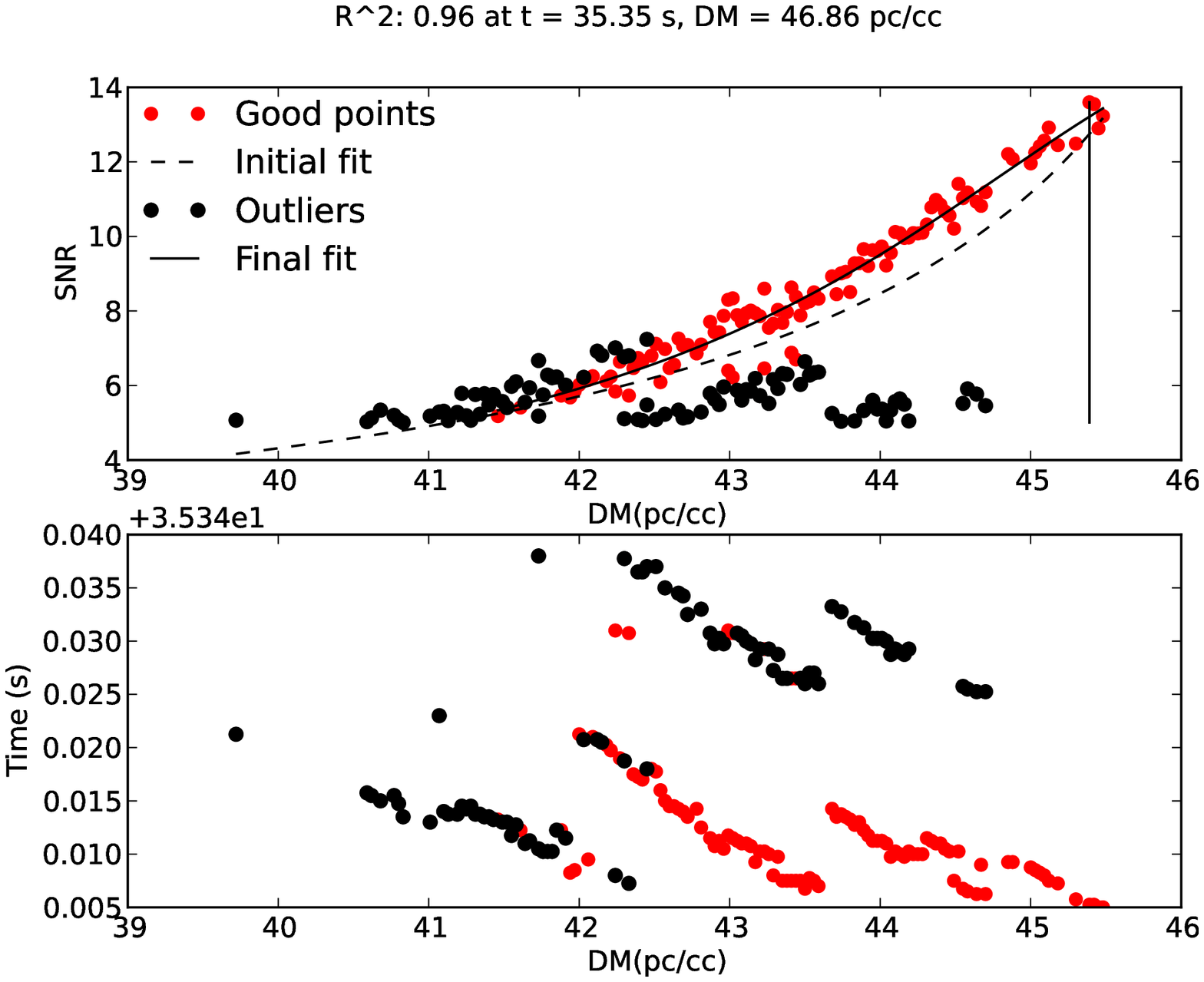}
\caption{Clusterrank fit resulting in the discovery of PSR~J0630+19. Top: $SNR$ vs. DM of the cluster with initial fit and final fit after removal of outliers. Bottom: the highly irregular structure of the cluster in DM-time space. In this case, the two shoulders of the $SNR$ vs. DM signature of the pulse resulted in two separate clusters of events. \label{fig_0630pulse1}}
\end{center}
\end{figure}

\begin{figure}[t]
\begin{center}
\includegraphics[width=\textwidth]{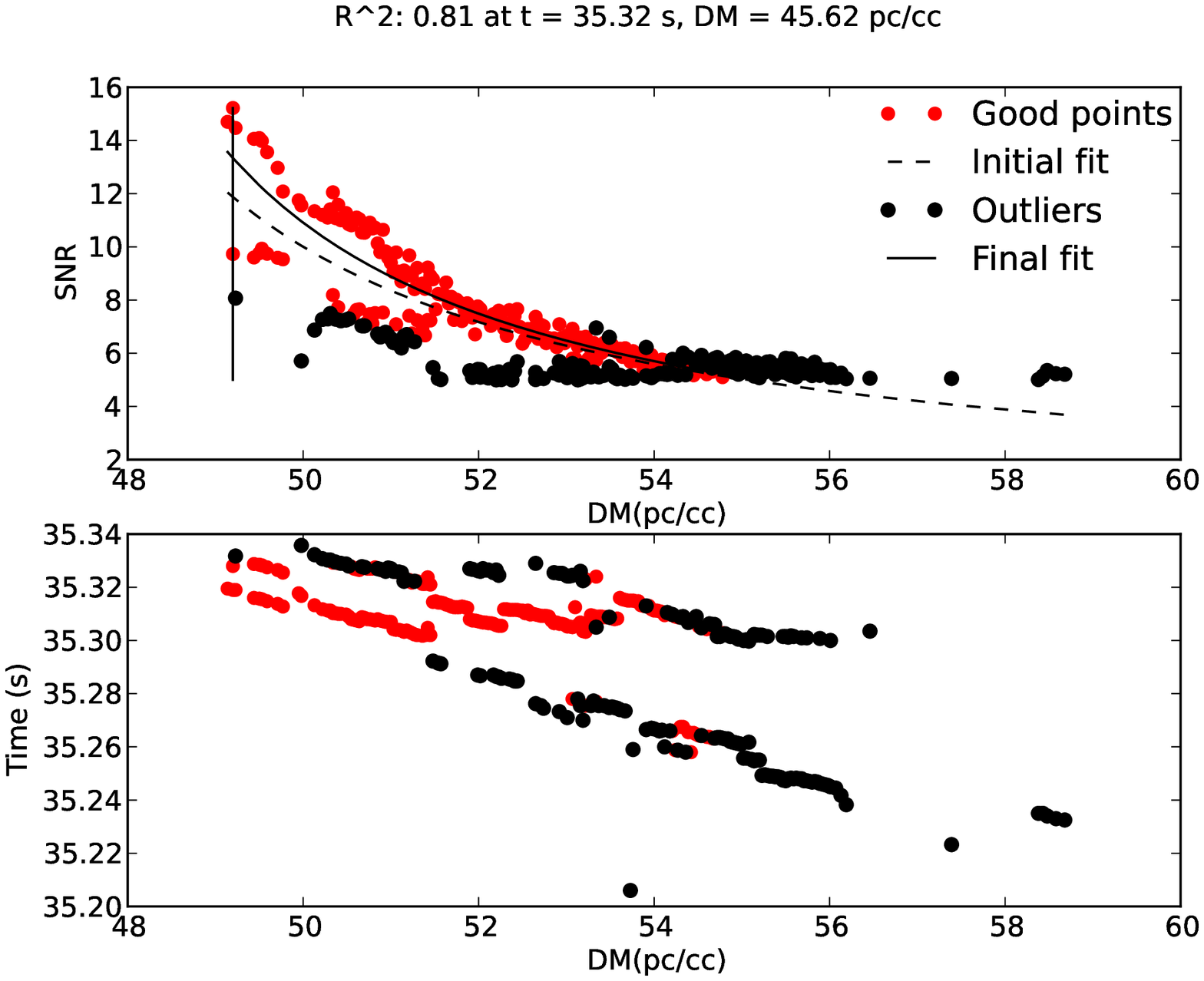}
\caption{Clusterrank fit resulting in the discovery of PSR~J0630+19. Top: $SNR$ vs. DM of the cluster with initial fit and final fit after removal of outliers. Bottom: the highly irregular structure of the cluster in DM-time space. In this case, the two shoulders of the $SNR$ vs. DM signature of the pulse resulted in two separate clusters of events. \label{fig_0630pulse2}}
\end{center}
\end{figure}

We note that the score is independent of the DM span of the cluster or the magnitude of event $SNR$s in the cluster. A weak pulse conforming well to Eqn.~\ref{eqn_lsqfit} will have a better score than a bright pulse that does not. 
%On the other hand, as a known pulsar drifts out of the beam, the observed $SNR$ of its pulses decreases and their shape in $SNR$ vs. DM space becomes less well-defined. The latter results in the score of pulses decreasing as the pulsar exits the beam. 
The score is also independent of the number of events in the cluster, as long as it is above the minimum required for the cluster to be fitted. Nor does the score depend on a cluster containing all events generated by the same pulse, as long as the cluster is well fitted by Eqn.~\ref{eqn_lsqfit}. This is the most significant difference between Clusterrank and codes like RRATtrap \citep{Karako15}, which {rely on the event with the highest SNR in a cluster to be present near the middle of the DM span of the cluster}. Figures \ref{fig_0630spplot}, \ref{fig_0630pulse1}, and \ref{fig_0630pulse2} show the discovery of PSR~J0630+19 made via fitting of two separate clusters corresponding to the two shoulders of a pulse in $SNR$ vs. DM space. 

The PRESTO single-pulse search, which constructs the event lists that serve as input to Clusterrank, by default does not search blocks in the dedispersed time series containing very bright, broad pulses typical of RFI. This approach is very effective in reducing the number of recorded events due to terrestrial sources, which can be overwhelming in some beams. However, this RFI excision scheme sometimes has an unintended effect on bright astrophysical pulses such that the resulting signature in $SNR$ vs. DM space is two shoulders with a missing peak in-between. Unlike RRATtrap, the ability of Clusterrank to detect dispersed pulses is unaffected by this, {even if the gap in DM is large enough that the two shoulders are processed as separate clusters.}

\subsection{RFI Rejection}\label{sec_rfi}

\begin{figure}[t]
\begin{center}
\includegraphics[width=\textwidth]{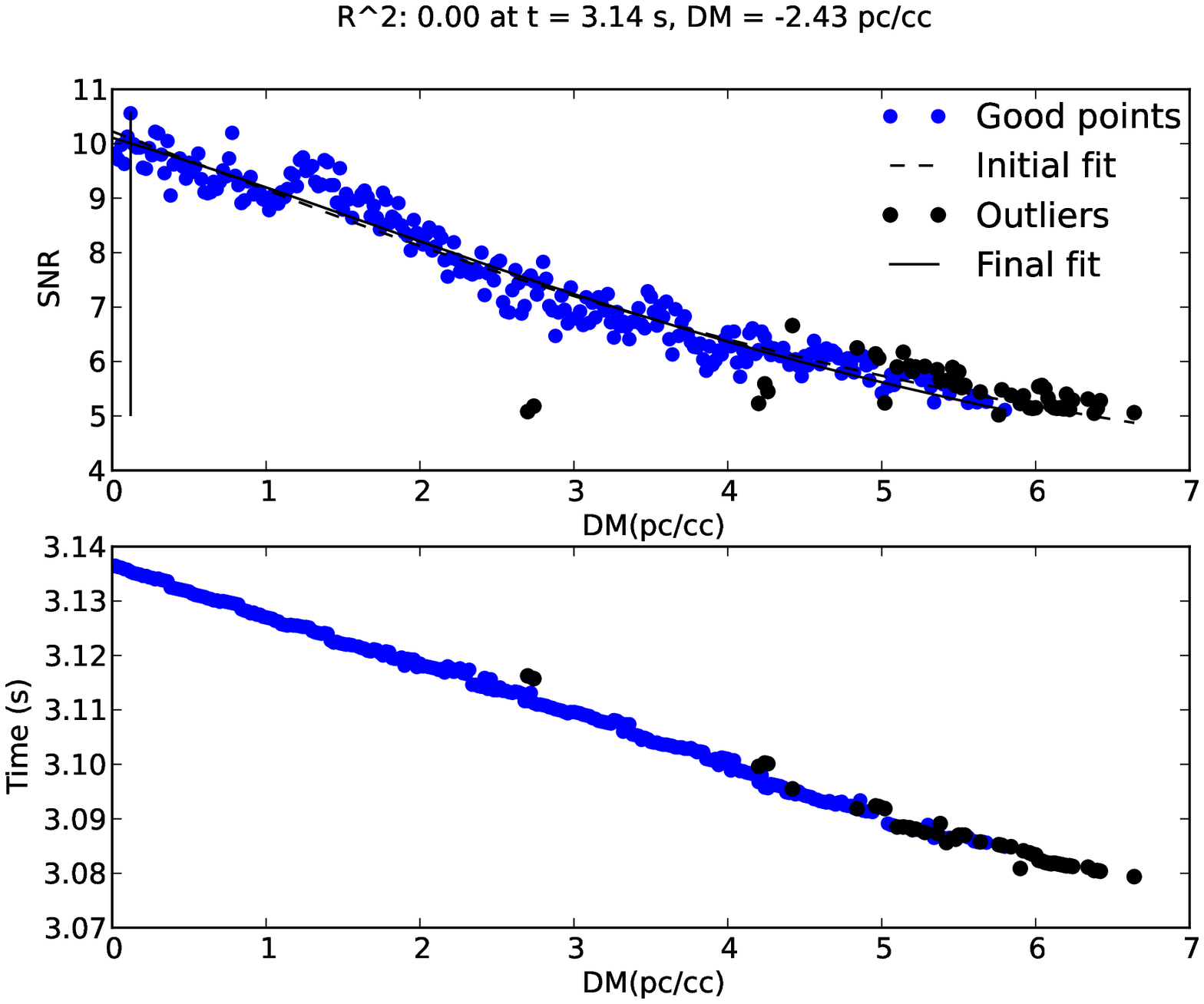}
\caption{Clusterrank fit of a cluster due to terrestrial RFI. Top: $SNR$ vs. DM of the cluster with initial fit and final fit after removal of outliers. Bottom: the structure of the cluster in DM-time space. The score of this cluster calculated from the fit is 0.95 and would have caused the single-pulse search plots for this beam to be selected for human inspection. However, the best-fit DM is negative, a non-physical result, and the score is set to zero. \label{fig_rfipulse1}}
\end{center}
\end{figure}

\begin{figure}[t]
\begin{center}
\includegraphics[width=\textwidth]{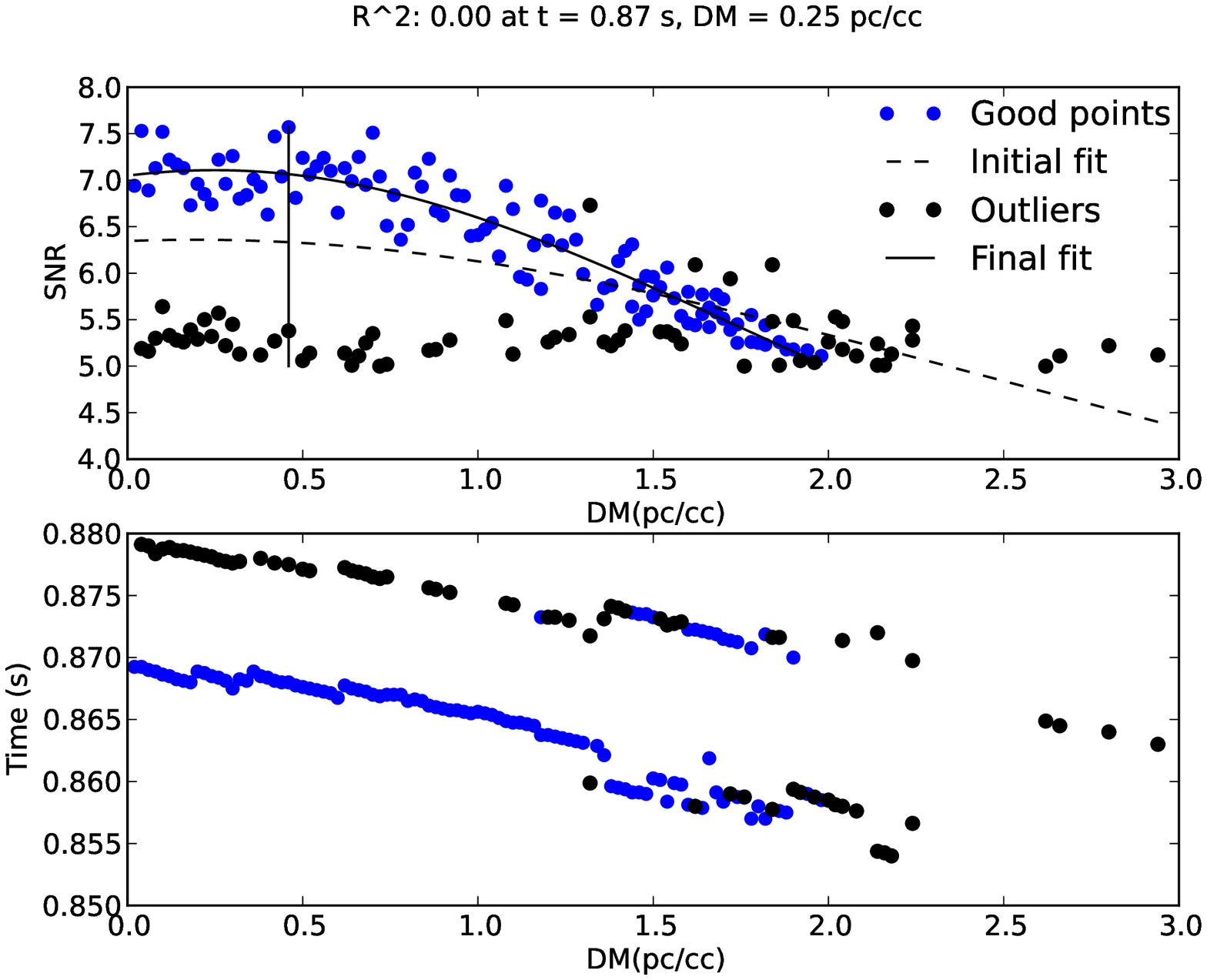}
\caption{Clusterrank fit of a cluster due to terrestrial RFI. Top: $SNR$ vs. DM of the cluster with initial fit and final fit after removal of outliers. Bottom: the structure of the cluster in DM-time space. The score of this cluster calculated from the fit is 0.87 and would have caused the single-pulse search plots for this beam to be selected for human inspection. However, the best-fit DM is $< 1$~pc~cm$^{-3}$ and the score is set to zero. \label{fig_rfipulse2}}
\end{center}
\end{figure}

PRESTO attempts to identify and remove RFI before making the event lists that Clusterrank operates on. Narrow-band and impulsive non-dispersed wideband signals are identified in the raw data and a time-frequency mask is constructed by PRESTO's tool {\tt rfifind}. During dedispersion, values of data points covered by the mask are replaced by a local average for that frequency channel. The PRESTO {\tt single\_pulse\_search.py} ignores blocks in the dedispersed time series containing bright, broad pulses, as described above. However, even after these RFI excision steps, there is still a significant number of events due to RFI in many of the event lists that serve as input to Clusterrank. We identify RFI in several ways. First, if the final fit to a cluster yields a negative best-fit DM or $W_{\rm ms}$, the score for that cluster is set to zero. Second, if a cluster is not fit by a negative DM or $W_{\rm ms}$ but the best-fit DM is less than 1~pc~cm$^{-3}$, the score for that cluster is set to zero. Figures \ref{fig_rfipulse1} and \ref{fig_rfipulse2} show two typical clusters with best-fit scores of 0.96 and 0.87 which are identified as RFI by these conditions. 

A different problem is presented by beams that are so contaminated by RFI that there are tens to hundreds of clusters in one of the four DM ranges considered by Clusterrank. Since each cluster fit is essentially an independent hypothesis test, a large number of tests done for events in the same DM range means that the likelihood of at least one false positive (RFI cluster with score $> 0.8$) for that beam is high. In order to mitigate this, we use a modification of the Bonferroni correction to the familywise error rate \citep{Bonferroni36}. We divide the best cluster score for each of the four DM ranges considered per beam by the base-10 logarithm of the number of clusters in that DM range rounded to the nearest integer. This means that 32 or more clusters in a single DM range would trigger the correction for that range. A known pulsar with 32 or more pulses within the AO327 integration time of one minute that are moreover bright enough to be detected in a single-pulse search would be detected in our concurrent periodicity search. {RFI pulses are often bright enough to cover a large range of DMs, and one genuine dispersed pulse plotted alongside 32 or more RFI pulses may be difficult to distinguish visually even if all single pulse plots were subjected to human inspection.} Since Clusterrank is geared towards detecting individual pulses, we consider the Bonferroni correction a good tradeoff for identifying this type of RFI contamination and excluding plots suffering from it from human inspection. {On average, 2.7\% of plots originally had $R^2 > 0.8$ but were excluded from human inspection after the Bonferroni correction was applied to their scores.}

\subsection{Performance}

In order to evaluate the performance of Clusterrank, we need to estimate the false positive and false negative rates, as well as what fraction of the total single-pulse search plots are selected for human inspection. Table~\ref{tab_performance} shows that overall for the Mock portion of AO327, 1.2\% of plots have score $> 0.9$ and 5.9\% have $0.8 < {\rm score} < 0.9$. However, the percentage of plots with score $> 0.9$ decreases from 1.9\% to 1.1\% between 2010 and 2011, and holds at 1.1\% for $2011 - 2013$. This is due to the fact that in early 2011, two sources of RFI in the 327~MHz band were identified on-site at Arecibo: cameras inside the Gregorian dome, and the rotation motors of the ALFA multibeam receiver. Subsequently, these were always disabled before the start of AO327 observing sessions. This decrease is evident for plots with scores of $0.8 - 0.9$ as well. 

Table~\ref{tab_performance} also shows what percentage of plots with score $> 0.8$ contain a detection of a known or new pulsar or RRAT. While the fraction of plots with a high score is driven by RFI, the fraction of high-ranked plots containing a detection is highly dependent on what part of the sky AO327 was observing in any year. As a low-frequency drift survey, AO327 does not typically get observing time near the oversubscribed inner Galactic plane. However, AO327 is often the only project that can take advantage of the telescope when it must be stationary for repairs or maintenance. During such times AO327 accumulates data and known pulsar detections in that region. The increased rate of detections in 2013 can be attributed to a lengthy painting job at the telescope platform during daytime in January - March 2013, coinciding with the time when the inner Galactic plane is above the horizon at Arecibo.

\subsubsection{False Positives}

{While the ideal way to compare the false positive rates of Clusterrank and the RRATtrap code of \cite{Karako15} is to do it per pulse, the published false positive rate of RRATtrap is on a per-plot basis. In addition, the efficiency of both codes in reducing the number of plots for human inspection is based on a per-plot score. Therefore, we proceed by comparing per-plot rates between the two codes.} 

RRATtrap scores 10\% of plots as ``excellent'' and selects them for human inspection. Ninety per cent of these plots contain false positives, resulting in an overall false positive rate of 9\%. If we consider Clusterrank plots with score $> 0.9$, the corresponding false positive rate is 1\% (Table~\ref{tab_performance}). If we consider Clusterrank plots with score $> 0.8$, the overall false positive rate is 7\%. Clusterrank results are from AO327 and RRATtrap results are from GBNCC and GBT350. The RFI environment is more challenging at Arecibo, where the radio-quiet zone around the telescope is smaller. The fact that Clusterrank has a smaller false positive rate despite that means that it is very effective in distinguishing RFI from astrophysical pulses. 

While 10\% of excellent RRATtrap plots contain a detection, 5\% of Clusterrank plots with score $> 0.9$ do. {The AO327 effective integration time and beam area area are 1 minute and 0.049~deg$^2$, respectively. The integration times of GBNCC and GBT350 are 2 and 2.3 minutes, respectively, while the beam area is 0.28~deg$^2$ for both GBT surveys. The remaining factor is the volume per unit solid angle searched, which depends on telescope sensitivity. Adapting the survey volume comparison in \cite{Deneva13} for single-pulse detections by using Eqn.~\ref{eqn_smin}, we find that for a 5~ms pulse $V_{\rm AO327,Mock}/V_{\rm GBNCC} \approx 4.5$ and $V_{\rm AO327,Mock}/V_{\rm GBT350} \approx 7.5$. Assuming that the 10\% RRATtrap detection rate is the same for GBT350 and GBNCC data and normalizing by the product of beam area, integration time, and volume per unit solid angle, we find that Clusterrank makes one detection in 1.14 times the volume per RRATtrap detection in GBT350 data, and in 0.78 times the volume per RRATtrap detection in GBNCC data.}

\begin{table}[t]
\begin{center}
\caption{Percentages of single-pulse search plots from the Mock portion of AO327 with $R^2 \ge 0.8$, the condition for human inspection. Percentages of each of these sets that contain a detection of a known or new pulsar or RRAT. \label{tab_performance}}
\begin{tabular}{lccccc}
\hline
Year & 2010 & 2011 & 2012 & 2013 & Overall \\
     & (\%) & (\%) & (\%) & (\%) & (\%) \\
\hline
Plots with $R^2 > 0.9$: & 1.9 & 1.1 & 1.1 & 1.1 & 1.2 \\
Plots with $0.8 \leq R^2 \leq 0.9$: & 7.3 & 5.5 & 6.1 & 5.4 & 5.9 \\
\\
\% of $R^2 > 0.9$ plots with detection: & 1.6 & 4.6 & 3.1 & 8.5 & 4.9 \\
\% of $0.8 \leq R^2 \leq 0.9$ plots with detection: & 0.6 & 0.6 & 0.3 & 0.5 & 0.4\\
\hline
\end{tabular}
\end{center}
\end{table}

%From Mock SP .singlepulse files processed with Clusterrank
% 2013: 
% Total plots: 153204
% 0.9*  plots: 1751 (1.1%); containing pulsars: 148 (8.5% of 0.9* plots)
% 0.8*  plots: 8219 (5.4%); containing pulsars: 43 (0.5% of 0.8* plots)
% 2012: 
% Total plots: 165563
% 0.9*  plots: 1801 (1.1%); containing pulsars: 57 (3.1% of 0.9* plots)
% 0.8*  plots: 10139 (6.1%); containing pulsars: 28 (0.3% of 0.8* plots)
% 2011: 
% Total plots: 41872 
% 0.9*  plots: 459 (1.1%); containing pulsars: 21 (4.6% of 0.9* plots)
% 0.8*  plots: 2312 (5.5%); containing pulsars: 13 (0.6% of 0.8* plots)
% 2010: 
% Total plots: 59432
% 0.9*  plots: 1172 (1.9%); containing pulsars: 30 (1.6% of 0.9* plots)
% 0.8*  plots: 4319 (7.3%); containing pulsars: 27 (0.6% of 0.8* plots)
% Totals:
% Total plots: 420071 (~934 h of data)
% 0.9*  plots: 5183 (1.2%); containing pulsars: 256 (4.9% of 0.9* plots)
% 0.8*  plots: 24989 (5.9%); containing pulsars: 111 (0.4% of 0.8* plots)

\subsubsection{False Negatives}

Clusterrank can produce three types of false negatives. Two are at the level of individual astrophysical pulses: (1) a pulse resulting in a cluster with $< 50$ events which is not fitted and (2) a pulse fitted with a score $< 0.8$. The third type of false negative is due to the Bonferroni correction described in Section~\ref{sec_rfi} and is at the level of the best cluster score recorded per DM range per beam which determines whether the respective plot is selected for human inspection. {Precisely determining the rate for the latter type of false negative would require inspecting the plots that triggered the Bonferroni correction, which comprise $\sim 20\%$ of all plots. By inspecting a random subset of these plots, we estimate that 0.02\% of all plots contain astrophysical pulses with $R^2 > 0.8$ but the best cluster score recorded for the plot was decreased to $< 0.8$ due to the Bonferroni correction. The false negatives in the inspected subset of plots were known pulsars whose high number of pulses within the beam triggered the correction.}

 Determining the rates for the first two types of false negatives precisely is not possible without visually inspecting all single-pulse search plots, which is what Clusterrank allows us to avoid. However, from results for a random set of beams containing known and new pulsar and RRAT detections we calculate that 2\% of astrophysical pulses result in clusters with $< 50$ events, which are not fitted by Clusterrank, and 27\% of astrophysical pulses have a best fit with $R^2 < 0.8$, which in the absence of other pulses would not select the plot for visual inspection. Using the same method, \cite{Karako15} estimate that $20\%$ of astrophysical pulses are not scored as ``excellent'' (but may still be marked as ``good'') by RRATtrap, at the expense of also producing more false positives than Clusterrank. 

We note that in the case of known pulsars, Clusterrank false negatives tend to occur as the pulsar enters and exits the beam, or if it traverses only the edge of the beam. As the pulsar moves away from the beam center, pulse $SNR$ decreases. While $R^2$ does not directly depend on $SNR$, the $SNR$ vs. DM shape that Clusterrank is fitting gradually becomes less pronounced and the pulse is detected at fewer trial DMs. 

\subsubsection{FRB Considerations}\label{sec_frbcons}

\begin{figure}[t]
\begin{center}
\includegraphics[width=\textwidth]{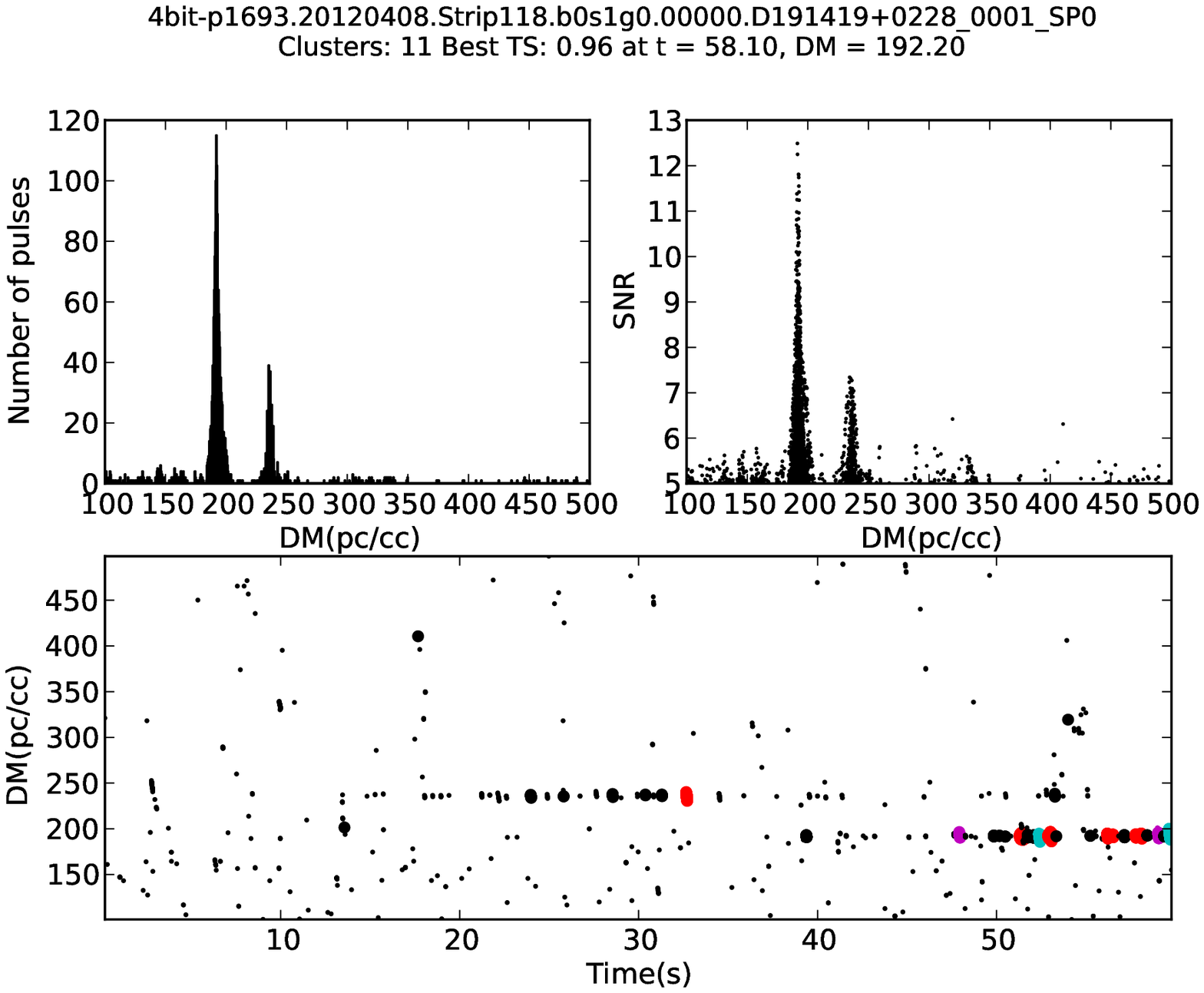}
\caption{Detection of the known pulsars J1914$+$0219 (DM = 233.8~pc~cm$^{-3}$) and J1915$+$0227 (DM = 192.6~pc~cm$^{-3}$) in the same beam. Most pulses are detected at very few DMs and result in clusters with $< 50$ events, which excludes them from being fitted by Clusterrank. However, each of these pulses taken in isolation is difficult to distinguish from noise clusters elsewhere in the DM vs. time panel, either visually or algorithmically. {Events belonging to clusters identified by Clusterrank are shown in red if the cluster $R^2 > 0.8$, magenta if $0.7 < R^2 \leq 0.8$, cyan if $0.6 < R^2 \leq 0.7$, green if $0.5 < R^2 \leq 0.6$, and blue if $R^2 \leq 0.5$. (There are no clusters with $0.5 < R^2 \leq 0.6$ or $R^2 \leq 0.5$ in this case.)} \label{fig_double}}
\end{center}
\end{figure}

An isolated highly dispersed pulse may be very difficult to distinguish from a noise cluster of events either algorithmically or visually. Pulsar surveys typically use trial DM lists with the interval between successive DMs increasing as the trial DM value increases (Table~\ref{tab_dedisp}). This is done to maximize computing efficiency: the detectability of pulsars with high DMs is limited by uncorrectable scattering broadening, not dispersion. However, it also means that an isolated, highly dispersed FRB pulse that is not bright enough to be detected at a large range of widely spaced DMs would be difficult to impossible to identify visually or by algorithms like Clusterrank and RRATRap, which rely on the $SNR$ vs. DM shape of the pulse. Figure~\ref{fig_double} shows a simultaneous detection of the known pulsars J1914$+$0219 (DM = 233.8~pc~cm$^{-3}$) and J1915$+$0227 (DM = 192.6~pc~cm$^{-3}$). Most of the two pulsars' pulses are not recognized as clusters and would be fitted poorly because they are detected at too few DMs.  In isolation, each of those pulses would be difficult to distinguish from clumps of noise events elsewhere on the plot. 

The spacing between successive trial DMs in the scattering-limited detection regime is typically informed by the fit of scattering time vs. DM made by \cite{Bhat04}, which is based on observations of Galactic sources. Unlike Galactic pulsars with DM~$\gtrsim 500$~pc~cm$^{-3}$, FRBs exhibit little to no scattering at 1.4~GHz. This can be explained by the fact that for FRBs, which are seen outside of the plane of our Galaxy, the bulk of the scattering material is in the host galaxy. For a scattering screen of the same size, the subtended angle as seen from Earth would be much smaller for the extragalactic source, essentially at the limit of the scattering screen being a point source. Therefore the difference in travel time for unscattered photons vs. photons scattered by the edges of the screen would be much smaller for an extragalactic than for a Galactic source. Correspondingly, the exponential scattering tail of the observed pulse caused by the spread of photon travel times due to scattering would be less prominent or absent for the extragalactic source. For these reasons, in order to maximize the chance of detecting highly dispersed, non-repeating FRBs, surveys should deliberately oversample the DM search space at high DMs.

\subsubsection{Clusterrank at High Frequencies}

In order to evaluate how well Clusterrank performs on data taken at a higher frequency commonly used in pulsar searching, we located PRESTO single-pulse search output files for the discovery observations of eight RRATs\footnote{\tt http://www2.naic.edu/\~{}palfa/newpulsars} and one FRB \citep{Spitler14} discovered by the PALFA survey at 1.4~GHz. {They were found by human inspection alone or facilitated by RRATtrap.} We ran Clusterrank on each set of files with no change in the algorithm or parameter values described above while appropriately specifying the PALFA observing frequency and bandwidth. The FRB received a score of 0.94. {Six of the RRATs received scores of $0.88 - 0.99$}. One RRAT was a false negative: its sole pulse resulted in fewer than 50 recorded events and therefore it was not fitted. The latter was the only data set from the older WAPP backend, which was used by PALFA until 2009. The lower sensitivity of WAPP vs. Mock PALFA observations means that there are fewer DMs, with larger spacings, in the trial DM list used to process WAPP data, than is the case for Mock data. Therefore, single-pulse search output from WAPP data has fewer events per pulse on average, and in that case an adjustment in the minimum number of events per cluster would improve the performance of Clusterrank. 

{The observation of one RRAT was severely contaminated by RFI, resulting in $> 1000$ clusters in the DM range containing the RRAT pulse. While the RRAT pulse received a score of 0.89, some RFI clusters received a score of 0.99. The high number of RFI clusters triggered the Bonferroni correction step in our algorithm (Section~\ref{sec_rfi}), which yielded an overall score of 0.33 for the DM range containing the RRAT pulse.} RFI in the PALFA bandwidth is dominated by several radars emitting pulses chirped at a variable rate. We find that in this situation the ability of Clusterrank to identify dispersed pulses based only on their shoulder shape at 327~MHz (Section~\ref{sec_ts}) becomes a liability at 1400~MHz. This is due to the fact that for the same DM range and pulse width, this shape becomes more linear with increasing frequency and therefore less likely to be uniquely identified with Eqn.~\ref{eqn_lsqfit} (\citealt{Cordes03}, Figure~4). This can be remedied by rejecting pulses whose best-fit DM is outside the DM range spanned by the cluster. 

%--Relate false positive rate to search volume ratios between AO327 and GBNCC/GBT350. Factors: beam size, beam transit time, search volume. Ideally, a measure of comparison would be something like false positives per unit search volume per hour. 

%\section{Histrank}

\section{New Pulsars}\label{sec_psrs}

\begin{figure}[t]
\begin{center}
\includegraphics{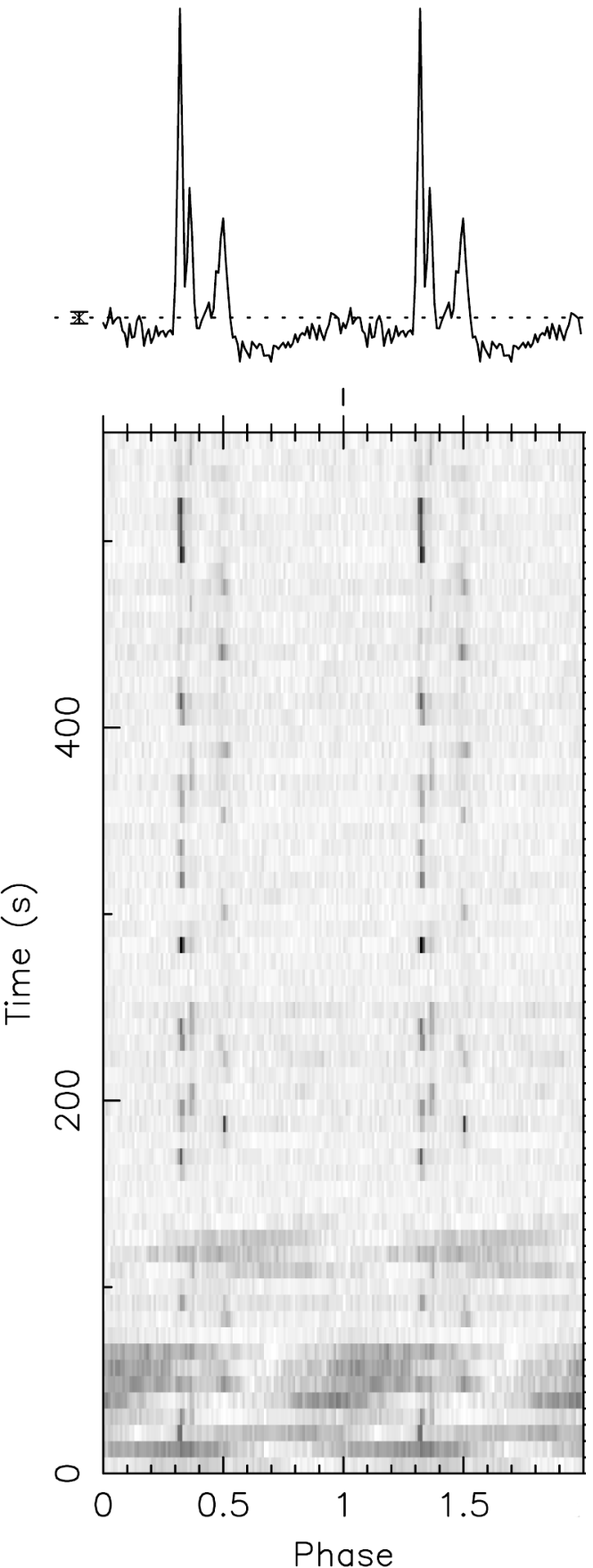}
\caption{Average pulse profile (top) and subintegration vs. pulse phase (bottom) for a confirmation observation of PSR~J1941+01 at 327~MHz. The pulsar switches between two modes with distinct pulse profiles at different phases. The dark areas across most of the period at $t < 150$~s are due to RFI.\label{fig_1941}}
\end{center}
\end{figure}

Clusterrank has facilitated the discovery of 22 new objects to date. Confirmation observations for all candidates use the 327~MHz receiver with the PUPPI backend and a $5 - 10$ minute integration time. Each confirmation observation is dedispersed with a range of DMs corresponding to the range over which pulses were detected in the discovery. If a period can be derived from single pulses either in the discovery or confirmation, the time series at the DM for which the pulse $SNR$ peaks is searched for periodic emission within a narrow range around that period. If a period cannot be derived or if the narrow search does not detect periodic emission, we perform a blind acceleration search of the time series. Periodicity searches of confirmation observations revealed that 14 of our 22 single-pulse discoveries are long-period pulsars, and Table~\ref{tab_psrs} summarizes their properties. In order to derive the peak flux density, we use a sky temperature $T_{\rm sky} = 50$~K \citep{Haslam82}. The PUPPI backend provides a bandwidth of 68~MHz and 2816 channels. The receiver temperature $T_{\rm rec} = 115$~K and gain $G = 11$~K/Jy are the same as for Mock observations. The peak flux density is
\be
S_{\rm pk} = \frac{\left(T_{\rm rec} + T_{\rm sky}\right)~SNR_{\rm prof}}{G \left(N_{\rm pol}~\Delta\nu~T_{\rm obs}/N_{\rm bin}\right)^{1/2}},
\ee
where $N_{\rm bin}$ is the number of bins in the averaged pulse profile and $SNR_{\rm prof}$ is the peak $SNR$ of that profile. 

The periods of the new pulsars are in the range $1.2 - 5.0$~s, with an average of 2.2~s. Surveys using short integration times and Fast Fourier Transform (FFT) periodicity search algorithms select against slow pulsars because there are few pulses per observation {\citep{Lazarus15}}, slow pulsars typically have duty cycles on the order of only $1 - 5\%$ {\citep{kgm04}}, and the pulses occur within a phase window that may be significantly wider than the width of an individual pulse. A Fast Folding periodicity search is more effective than an FFT search when there are few rotation periods within an observation (\citealt{Staelin69}, \citealt{Kondratiev09}). We plan to reprocess AO327 survey data with a Fast Folding search, which will be sensitive to slow periodic emitters missed by the PRESTO FFT-based periodicity search that moreover do not emit pulses bright enough to be detected by a single-pulse search. Exhaustive searches for pulsars that are selected against by most widely used search algorithms are important for constructing a more complete picture of the period and age distribution of the Galactic pulsar population and relating these statistics to independent measures of pulsar formation such as the Galactic supernova rate. 

Two more narrowly defined subsets of slow pulsars that are selected against in FFT-based periodicity searches are also represented in Table~\ref{tab_psrs} and overrepresented among Clusterrank discoveries compared to the general pulsar population. PSRs~J1749+16 and J1750+07 null for tens of seconds at a time. The individual pulses of PSR~J1750+07 have a peak flux density of up to $\sim 150$~mJy and are bright enough to trigger PRESTO's ``bad block'' flagging (Section~\ref{sec_rfi}), yet its integrated profile peak flux density is only 15.5~mJy because its nulling fraction is $> 50\%$. 

PSR~J1941+01 has the highest DM of all AO327 discoveries to date, 133~pc~cm$^{-3}$. It is a mode-switching pulsar and alternates in a quasi-periodic manner between two states with distinct pulse shapes and phase windows (Figure~\ref{fig_1941}). In addition, one of the modes exhibits subpulse drifting. The emission of pulsars with similar properties has been explained by the carousel model, where emission sub-beams circulate around the magnetic field axis, giving rise to emission patterns that repeat on time scales of many pulse periods as the observer's line of sight crosses different sub-beam configurations (e.g. \citealt{Rankin08}). We are pursuing multi-frequency polarimetric observations of J1941+01 in order to map its emission region cone in altitude as well as cross-section and defer a more detailed analysis to a separate paper. 

%--Pav for the new pulsars: 2198.21
%--Pav for pulsars in ATNF with P>20ms: 0.929
%--Pav for RRATs from the RRATalog: 2.2826

\begin{table}[t]
\begin{center}
\caption{New pulsars discovered by AO327 via a single-pulse search. All objects were discovered via single-pulse search and identified by the Clusterrank code described in this paper. $R^2$ is the value for the highest-ranked pulse in the discovery Mock observation. Confirmation observations with the more sensitive PUPPI backend yielded periodic detections and many pulses for all objects. $W_{prof}$ is the full-width, half-maximum width of the folded pulse profile and $S_{pk}$ is the peak flux density derived from it. {$N_{p}$ is the number of pulses in the discovery observation.}
%$S_{pk}$ is the peak flux density of the brightest pulse in the discovery Mock observation. 
%{\bf (XXX Include columns for Spk of brightest pulse in discovery and Sav for folded profile?)}
\label{tab_psrs}}
\begin{tabular}{lccccccccc}
\hline
Name & RA  & DEC  & $P$ & DM  & $W_{prof}$ & $S_{pk}$ & $N_{p}$ & $R^2$   \\
     & (hh:mm:ss)$^a$ & (dd:mm) & (ms) & (pc~cm$^{-3}$) & (ms) & (mJy) &   \\
\hline
J0011+08 & 00:11:34 & 08:10 & 2552.87 & 24.9 & 28 & 12.3 & 7 & 0.91  \\ %9 pulses in disc
J0050+03 & 00:50:31 & 03:48 & 1366.56 & 26.5 & 33 & 15.2 & 7 & 0.87  \\ %8 pulses in disc
J0611+04 & 06:11:18 & 04:06 & 1674.43 & 69.9 & 81 & \phn3.5 & 2 & 0.94  \\ %2 pulses in disc, 64+2x600s+3x300s+1x900s+1x120s
J0630+19 & 06:30:04 & 19:37 & 1248.55 & 48.1 & 35 & \phn3.6 & 1 & 0.96 \\
J1656+00 & 16:56:41 & 00:26 & 1497.85 & 46.9 & 34 & 11.4 & 1 & 0.95  \\ %2 pulses in disc
J1738+04 & 17:38:25 & 04:20 & 1391.79 & 23.6 & 28 & 14.1 & 8 & 0.91  \\ %8 pulses, 64s (some consecutive)
J1743+05 & 17:43:16 & 05:29 & 1473.63 & 56.1 & 55 & \phn5.9 & 3 & 0.90 \\
J1749+16 & 17:49:29 & 16:24 & 2311.65 & 59.6 & 61 & \phn7.3 & 6 & 0.81 \\ % this R^2 was for an RFI cluster; find out what it was for the highest-ranked actual pulse
J1750+07 & 17:50:40 & 07:33 & 1908.81 & 55.4 & 60 & 15.5 & 3 & 0.94 \\
J1938+14 & 19:38:19 & 14:42 & 2902.51 & 74.2  & 95 & \phn5.2 & 4 & 0.85  \\ %4 pulses, 64s
J1941+01$^b$ & 19:41:58 & 01:46 & 1404.73 & 133.3 & 40 & 18.4 & 6 & 0.95  \\ %7 pulses in disc
J1946+14 & 19:46:52 & 14:42 & 2282.44 & 50.3  & 50 & 11.4 & 3 & 0.90  \\ %2 pulses in disc
J1956+07 & 19:56:35 & 07:16 & 5012.48 & 61.3 & 125 & \phn3.6 & 3,2$^c$ & 0.96 \\ %4+2 pulses in 2 blind detections 
J2105+07 & 21:05:27 & 07:57 & 3746.63 & 52.6  & 126 & 35.1 & 5 & 0.97  \\ %5 pulses in disc, 64s
\hline
\end{tabular}
\end{center}
$^a$ RA and DEC are given in the J2000 coordinate system. The uncertainties in both coordinates are 7.5\amin, the 327~MHz beam radius, unless otherwise indicated. \\
$^b$ J1941+01 is a mode-switching pulsar and exhibits two distinct pulse profiles corresponding to two modes. $W_{prof}$ and $S_{pk}$ given here refer to the state with the brighter peak. \\
$^c$ J1956+07 was identified by Clusterrank in two 1-minute data spans from observations taken on different days. 
\end{table}

\section{New RRATs}\label{sec_rrats}

Eight of the objects discovered with the help of Clusterrank do not exhibit periodic emission in follow-up observations and therefore we provisionally classify them as RRATs (Table~\ref{tab_rrats}). We were able to estimate the rotation periods for four RRATs based on the intervals between pulses detected within one observation. In the case of J1603+18, we detected three pulses emitted on consecutive rotations, separated by intervals of $\sim 0.503$~s. For the remaining four RRATs, the intervals between detected pulses are uneven and significantly longer, and therefore the estimated period may be an integer multiple of the actual rotation period. 

We calculate the peak flux density of the brightest pulse in the discovery observation of each RRAT from the radiometer equation for single pulses:
\be
S_{\rm pk} = \frac{\left(T_{\rm rec} + T_{\rm sky}\right)~SNR_{\rm pk}}{\left(N_{\rm pol}~\Delta\nu~W\right)^{1/2}}, 
\ee
where $SNR_{\rm pk}$ is the peak signal-to-noise ratio of the brightest pulse, and $W$ is its full width at half maximum. 

{RRAT candidate J0156+04 (Table~\ref{tab_rrats}) remains unconfirmed. Six confirmation attempts of 10 minutes each were made using the 327~MHz Arecibo receiver and the PUPPI backend. J0156+04 exhibits the typical signature of a dispersed pulse with one peak in $SNR$ vs. DM that is easily fitted by Clusterrank and recognized visually (Figure~\ref{fig_0156spplot}, Figure~\ref{fig_0156pulse1}). However, even the brighter of the two detected pulses is too weak for the dispersion sweep to be visible in a plot of the raw data in time-frequency space around the pulse arrival time.}

The properties of J0156+04, two or more pulses in close succession and consistent non-detections in multiple follow-up observations, are shared by a small subset of radio transients detected by virtually every pulsar survey using a single-pulse search. Two recent examples are J1928+15 \citep{Deneva09} and J1336$-$20 \citep{Karako15}. This type of transient emission may indicate an object that is dormant or not beamed toward the Earth and whose magnetosphere is perturbed sporadically by accretion of debris from an asteroid belt \citep{Cordes08}.

\begin{table}[t]
\begin{center}
\caption{New RRATs discovered by the AO327 drift survey. All objects were discovered via single-pulse search and identified by the Clusterrank code described in this paper. $R^2$ is the value for the highest-ranked pulse in the discovery Mock observation. $W$ is the full-width at half-maximum of the brightest detected pulse, and $S_{pk}$ is its peak flux density. {$N_{p}$ is the number of pulses in the discovery observation.} Also listed is the average pulse rate, defined as the ratio of the total number of pulses detected to the total observation time. For objects that have been detected in only one observation we take this to be an upper limit. The last column lists if an object has had a successful confirmation detection after the discovery. %$N_{det}/N_{obs}$ is the ratio of observations with at least one detected pulse vs. the total number of observations. 
\label{tab_rrats}}
\begin{tabular}{lcccccccccc}
\hline
Name & RA  & DEC  & P & DM  & $W$ & $S_{pk}$ & $N_{p}$ & $R^2$ & Rate & Conf\\ %& $N_{det}/N_{obs}$ \\
     & (hh:mm:ss)$^a$ & (dd:mm) & (ms) & (pc~cm$^{-3}$) & (ms) & (Jy) & & & (hr$^{-1}$) \\ %&  \\
\hline
J0156+04 & 01:56:01 & 04:02 & - & 27.5 & \phn3.8 & 0.3 & 2 & 0.97 & $\leq 2$ & \\ %& 1/6 \\ %2 pulses in disc, 64+6x600s
J0544+20 & 05:44:12 & 20:50 & - & 56.9 & \phn2.3 & 0.3 & 1 & 0.95 & 4 & Y \\ %up to MJD 57060: Tobs_tot = 14394s, Npulses_tot = 14, Nobs_detected = 11, Nobs_tot = 22
J0550+09 & 05:50:28 & 09:51 & 1745 & 86.6 & 22.5 & 0.1 & 3 & 0.93 & 47 & Y \\ %& 2/2 \\ %4 pulses, 64+240s
%J1017+02 & 10:17:17 & 02:19 & - & 22.5 & \phn0.5 & 1.9 & 2 & 0.83 & $\leq 0.7$ & \\ %& 1/4 \\ %2 pulses in disc, 64+2x600+900+480+2400s
J1433+00 & 14:33:30 & 00:28 & - & 23.5 & \phn3.8 & 0.3 & 1 & 0.94 & 2 & Y \\ %& 2/7 \\ %1 pulse in disc, 1 pulse in conf. 64+6x600s
J1554+18 & 15:54:17 & 18:04 & - & 24.0 & \phn7.6 & 0.2 & 1 & 0.89 & 11 & Y \\ %1 pulse in disc, 1 in confirmation, 64+600s
J1603+18 & 16:03:34 & 18:51 & \phn503 & 29.7 & \phn8.8 & 0.2 & 1 & 0.94 & 4 & Y \\ %up to MJD 57060: Tobs_tot = 4524s, Npulses_tot = 5, Nobs_detected = 2, Nobs_tot = 10
J1717+03 & 17:17:56 & 03:11 & 3901 & 25.6 & \phn8.4 & 0.2 & 1 & 0.91 & 8 & Y \\ %up to MJD 57060: Tobs_tot = 6154s, Npulses_tot = 23, Nobs_detected = 7, Nobs_tot = 12
J1720+00 & 17:20:55 & 00:40 & 3357 & 46.2 & \phn7.2 & 0.2 & 1 & 0.97 & 33 & Y \\ %up to MJD 57060: Tobs_tot = 4834s, Npulses_tot = 44, Nobs_detected = 11, Nobs_tot = 12
%J1907+06 & 19:07:42 & 06:09 & 172 & 24.0 & \phn0.5 & 1.7 & 4 & 0.77$^b$ & $\leq 4$ & \\ %& 1/6 \\ %4 pulses in disc, 64+6x600s (the 0.97 R^2 refers to RFI in the beam, check actual pulse R^2, it would still be >0.8)
\hline
\end{tabular}
\end{center}
$^a$ RA and DEC are given in the J2000 coordinate system. The uncertainties in both coordinates are 7.5\amin, the 327~MHz beam radius, unless otherwise indicated. 
\end{table}

\section{Perytons}\label{sec_per}

\begin{figure}[t]
\begin{center}
\includegraphics[width=\textwidth]{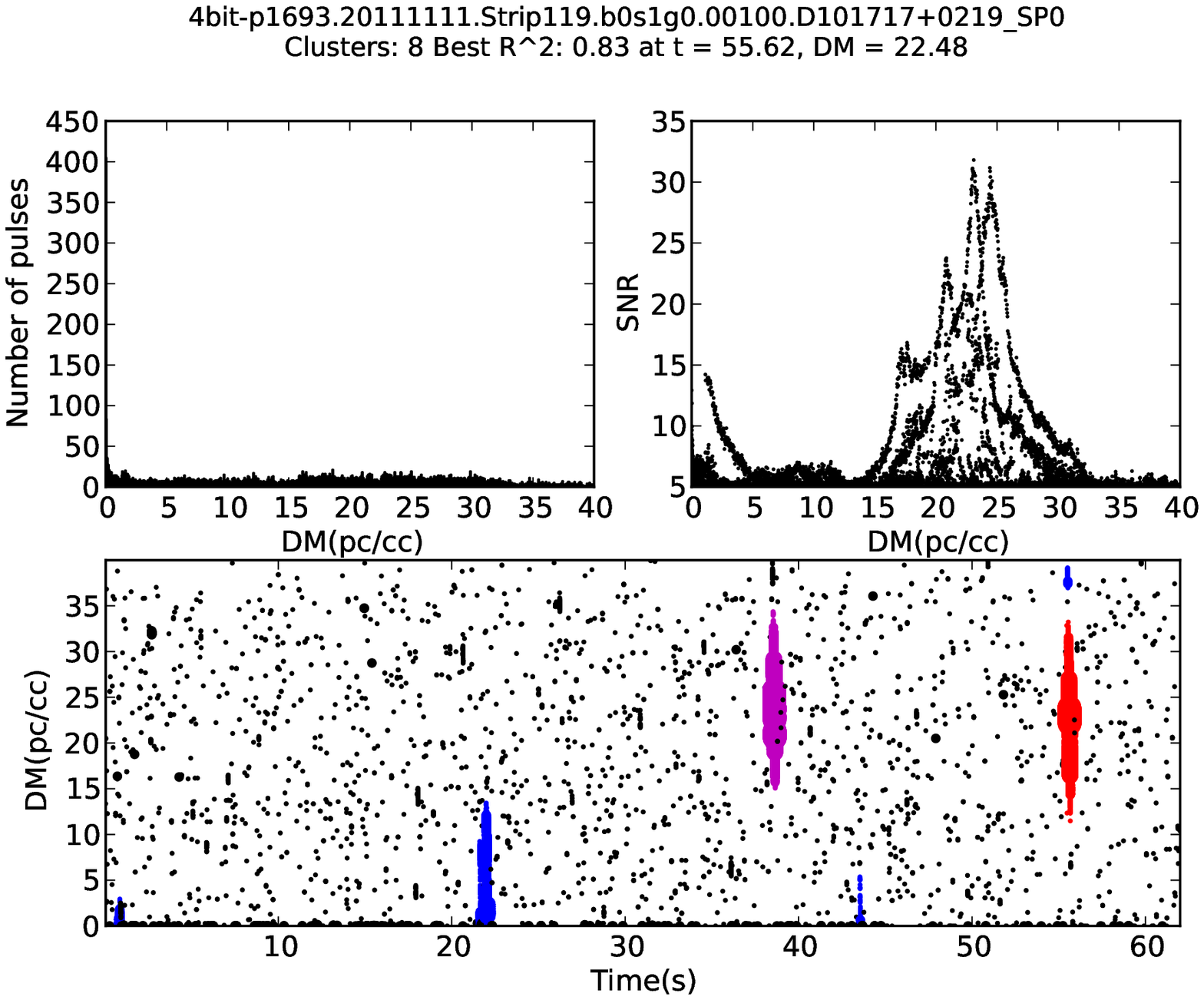}
\caption{Single-pulse search plot of the detection of peryton P1017+02. Top: Histograms of the number of events vs. DM (left) and event $SNR$ vs. DM (right). Bottom: Events are plotted vs. DM and time. Larger marker sizes correspond to higher $SNR$. Events belonging to clusters identified by Clusterrank are shown in red if the cluster $R^2 > 0.8$, magenta if $0.7 > R^2 > 0.8$, and blue if $R^2 < 0.5$. The multi-peaked DM vs. $SNR$ signatures of the two pulses at $t \sim 39$~s and $t \sim 56$~s present a challenge for Clusterrank and indicate a multi-peaked pulse profile. \label{fig_1017spplot}}
\end{center}
\end{figure}

\begin{figure}[t]
\begin{center}
\includegraphics[width=\textwidth]{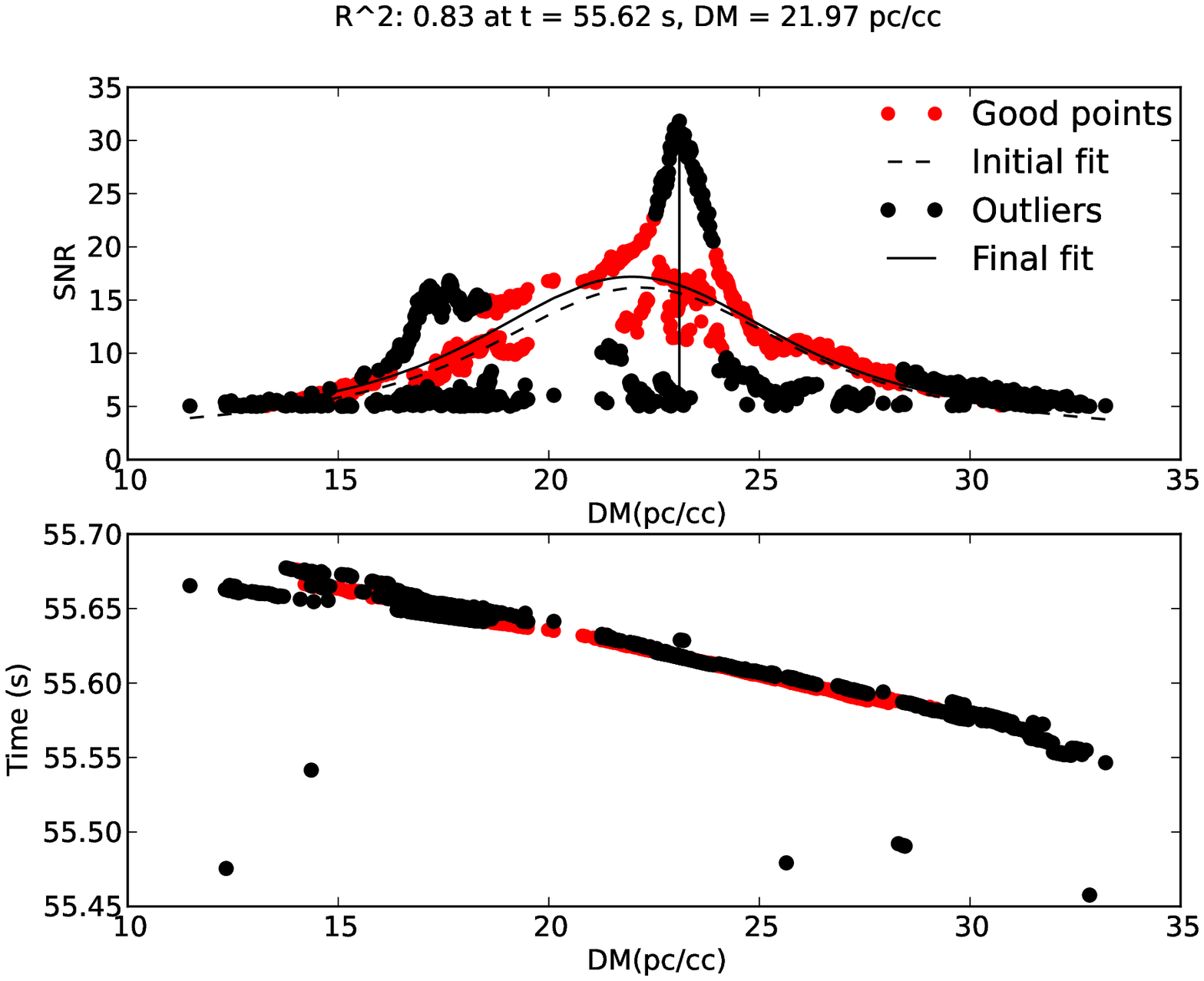}
\caption{Clusterrank fit for the pulse with the highest score in the observation of peryton P1017+02. Top: $SNR$ vs. DM for events in the cluster corresponding to this pulse, along with initial and final fits. Bottom: the structure of the cluster is shown in DM-time space. \label{fig_1017pulse}}
\end{center}
\end{figure}

\begin{figure}[t]
\begin{center}
\includegraphics[width=\textwidth]{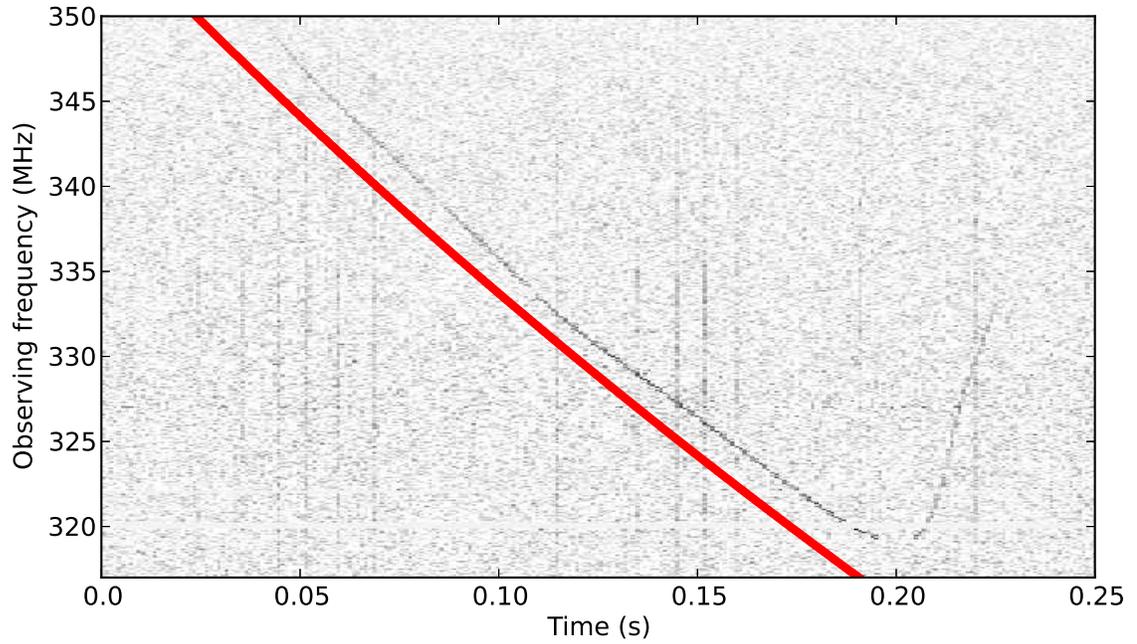}
\caption{A time-frequency plot of the raw data around the arrival time of the brighter pulse of peryton P1017+02 ($t \sim 56$~s in Figure~\ref{fig_1017spplot}; Figure~\ref{fig_1017pulse}). The red line shows the dispersion sweep for the best-fit DM of this pulse, 21.97~pc~cm$^{-3}$ according to the cold-plasma dispersion relation. Darker vertical lines are caused by wide-band, non-dispersed terrestrial RFI. \label{fig_peryton}}
\end{center}
\end{figure}

\begin{figure}[t]
\begin{center}
\includegraphics[width=\textwidth]{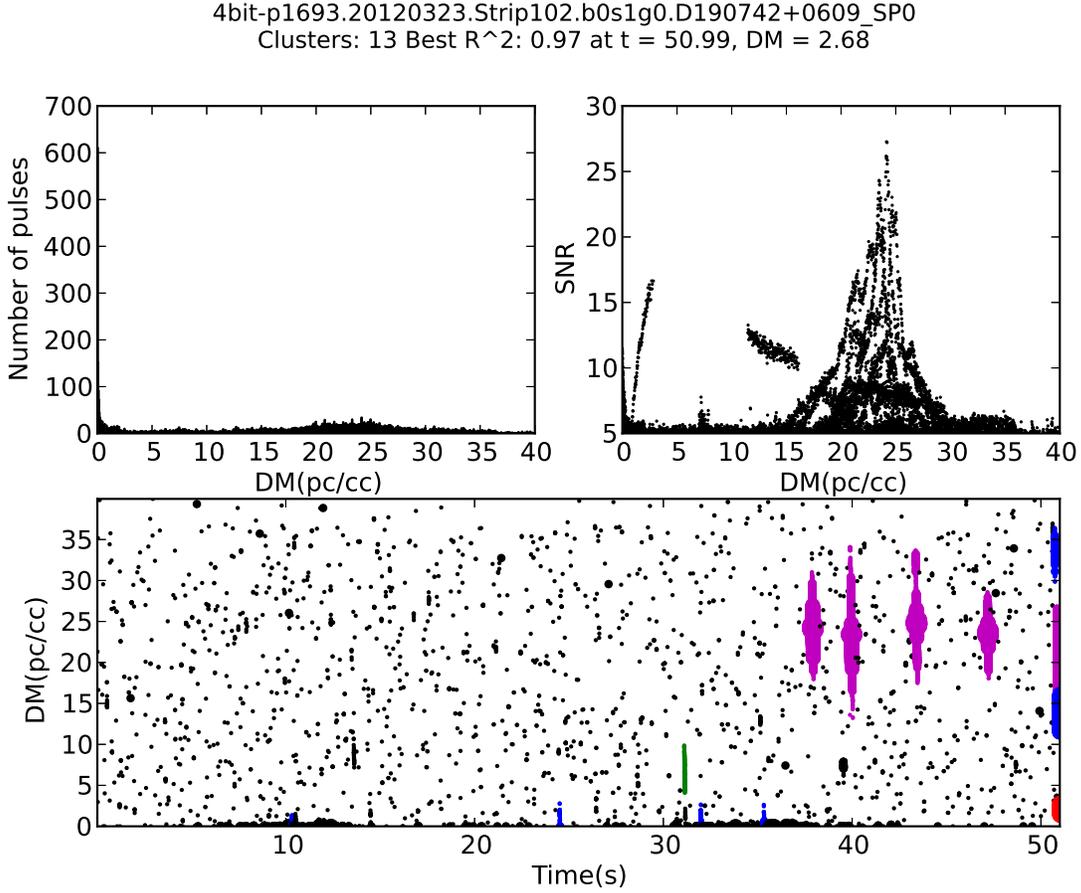}
\caption{Single-pulse search plot of the detection of peryton P1907+06. Top: Histograms of the number of events vs. DM (left) and event $SNR$ vs. DM (right). Bottom: Events are plotted vs. DM and time. Larger marker sizes correspond to higher $SNR$. Events belonging to clusters identified by Clusterrank are shown in red if the cluster $R^2 > 0.8$, magenta if $0.7 > R^2 > 0.8$, green if $0.6 > R^2 > 0.5$,and blue if $R^2 < 0.5$. The multi-peaked DM vs. $SNR$ signatures of the four pulses at $DM \sim 24$~pc~cm$^{-3}$ present a challenge for Clusterrank and indicate a multi-peaked pulse profile. \label{fig_1907spplot}}
\end{center}
\end{figure}

\begin{figure}[t]
\begin{center}
\includegraphics[width=\textwidth]{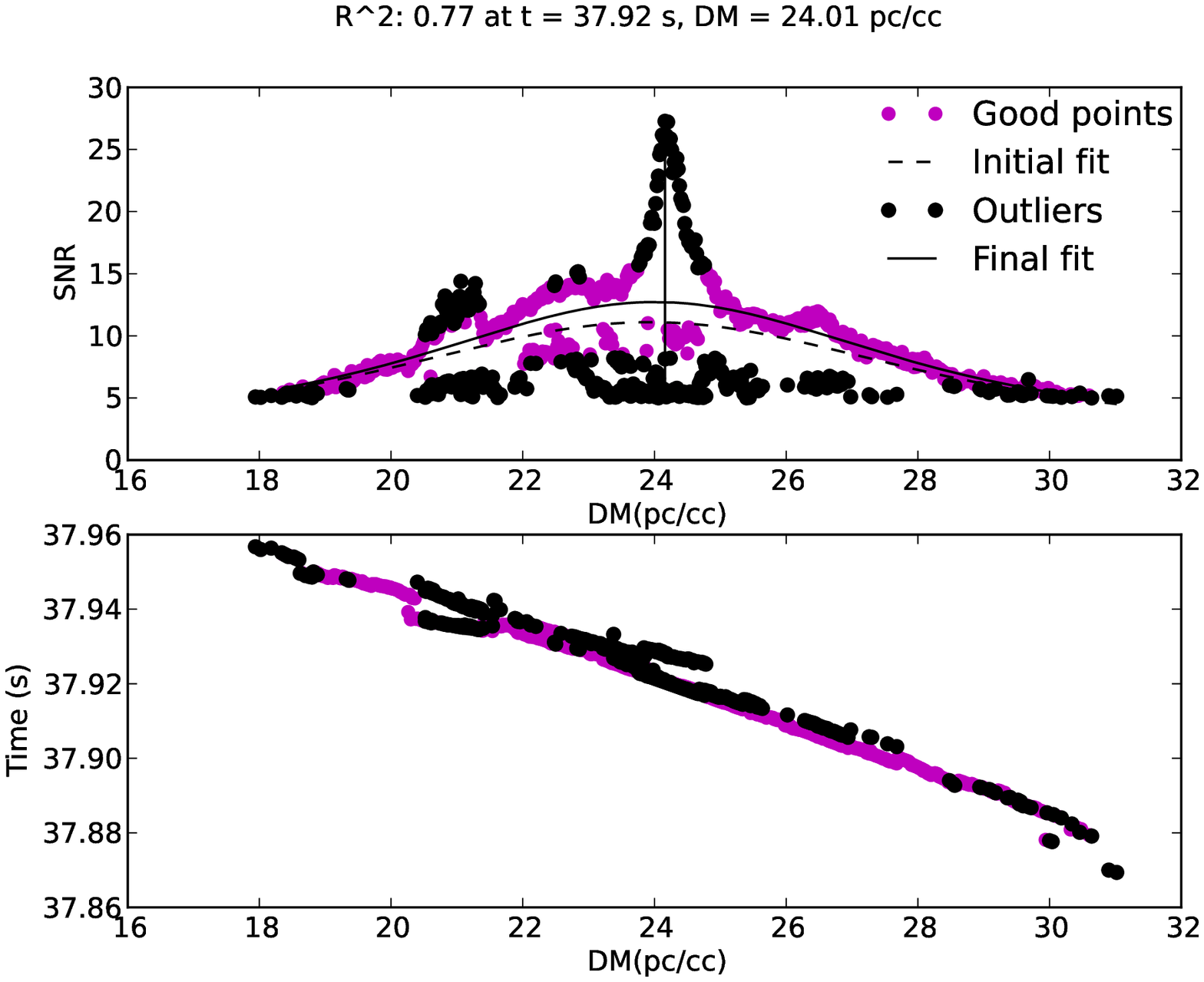}
\caption{Clusterrank fit for the pulse with the highest score in the 
observation of peryton P1907+06. Top: $SNR$ vs. DM for events in the cluster corresponding to this pulse, along with initial and final Clusterrank fits. Bottom: the structure of the cluster is shown in DM-time space. \label{fig_1907pulse}}
\end{center}
\end{figure}

{Two bright signals assigned high scores by Clusterrank were also not detected in follow-up observations. Further inspection revealed that their sweep in time-frequency space does not completely conform to the cold-plasma dispersion relation. We classify them as perytons, terrestrial RFI mimicking a dispersed signal \citep{Burke11}. 

The total follow-up observation time is one hour for peryton P1907+06, and 1.6~h for peryton P1017+02. P1017+02 (Figure~\ref{fig_1017spplot}, Figure~\ref{fig_1017pulse}) and P1907+06 (Figure~\ref{fig_1907spplot}, Figure~\ref{fig_1907pulse}) are qualitatively different from J0156+04, as well as from any other single-pulse source detected by AO327. They are readily recognized visually since the corresponding clusters of events have a limited extent in DM and a definite peak in $SNR$ vs. DM that is consistent between pulses from the same source. However, they exhibit complex substructure and secondary peaks in $SNR$ vs. DM space. This corresponds to a similarly complex resolved substructure in each pulse, such that different components are aligned and summed at different trial DMs. P1017+02 and P1907+06 also differ from the RRATs in Table~\ref{tab_rrats} in that they are brighter by an order of magnitude. 

Figure~\ref{fig_peryton} shows the time-frequency structure of the brightest peryton pulse, also shown in Figures~\ref{fig_1017spplot} and \ref{fig_1017pulse}. All pulses from P1017+02 and P1907+06 exhibit the same tickmark-like signature, indicating a chirped signal with an abrupt sign reversal of the chirp rate. Such signals have also been detected by the GBT350 survey (C.~Karako-Argaman, private communication\footnote{\tt http://www.physics.mcgill.ca/\~{}karakoc/waterfall\_0526-1908.png}) and anecdotal accounts suggest that they may be generated in the process of shutting down the transmitters of some aircraft.} 

\section{Distributions and Populations}\label{sec_stats}

{Following the analysis presented by \cite{Karako15}, we compare the properties of Clusterrank discoveries with those of known pulsars and RRATs.} The average period for pulsars found by Clusterrank is 2.2~s. This is larger than the 0.9~s average period for non-MSP ($P > 0.02$~s) pulsars from the ATNF catalog\footnote{\tt http://www.atnf.csiro.au/research/pulsar/psrcat}, but similar to the 2.3~s average period of RRATs with measured periods in the RRATalog\footnote{\tt http://astro.phys.wvu.edu/rratalog}. However, a Kolmogorov-Smirnov (K-S) two-sample test between Clusterrank pulsar discoveries and RRATalog objects yields a p-value of 0.017, suggesting that the two sets are not drawn from the same period distribution. {In contrast, the same type of test between Clusterrank RRAT discoveries with period estimates and RRATalog objects yields a p-value of 0.93, indicating consistency with the null hypothesis of the same underlying period distribution for both sets.} Of the 115 objects currently in the RRATalog, 85 were discovered by surveys operating at 1400~MHz and using significantly longer integration times than AO327. This suggests that neither the observing frequency nor the number of rotations within a standard survey observation result in strong selection effects when we consider the period distribution of RRAT discoveries. The latter may be explained by the fact that follow-up observations of newly discovered RRATs are typically longer than the discovery observation and thus slow pulsars discovered by single-pulse search are likely to be identified as periodic emitters when reobserved. The weak dependence on observing frequency is due to the fact that dispersion ($\propto \nu^{-2}$) and scattering broadening ($\propto \nu^{-4}$) do not significantly select against detecting RRATs at 327 vs. 1400~MHz due to their long periods. 

%{\bf The p-value for a K-S test between the period distributions of pulsars and RRATs discovered by Clusterrank is 0.73, indicating a possible overlap between RRATs and a subset of slow canonical pulsars. We note, however, that Clusterrank's sampling of canonical pulsars is biased by the Bonferroni correction, which would reject pulsars with bright pulses and periods $\lesssim 2$~s.} Pulsar spectra are typically steep, $S \propto \nu^\alpha$, where the average $\alpha = -1.7$. Some RRATs may be pulsars whose pulse intensity distribution has a long tail, such that only the brightest pulses are detected, at the discovery frequency or altogether. J0627+16, an RRAT discovered at 1400~MHz, was detected as a periodic emitter at 327~MHz \citep{Deneva09}. \cite{Weltevrede06} argue that if B0656+14 were farther away, it would be detectable only as an RRAT due to its long-tailed pulse intensity distribution. 

Since the DM and spatial distribution of a set of pulsar discoveries depends on what region of the sky survey observations are targeting, we can meaningfully compare Clusterrank discoveries only with AO327 periodicity search discoveries. Performing two-sample K-S tests between these two sets we obtain a p-value of 0.60 for their DM distributions, and p-values of 0.87 and 0.98 for the Galactic latitude and longitude, respectively. Therefore we can conclude that Clusterrank discoveries are drawn from the same spatial and DM distribution as AO327 periodicity search discoveries. 

\section{FRB Population Limits}\label{sec_frblimits}

We did not find any FRBs in the AO327-Mock data set presented in this paper. However, as we outline in Section~\ref{sec_frbcons}, the methods usually employed for calculating the optimal trial DM list for a pulsar search assume a relationship between dispersion and scattering typical of Galactic sources that does not hold for known FRBs. We plan to reprocess all AO327 survey data with a trial DM list optimized for detecting highly dispersed but not significantly scattered FRB pulses. 

The first upper limits on the all-sky FRB rate are from Parkes pulsar surveys and their FRB detections. For FRBs with a flux density $S \gtrsim 3$~Jy at 1.4~GHz, \cite{Thornton13} estimate a rate of $1.0^{+0.6}_{-0.5} \times 10^4$~sky$^{-1}$~day$^{-1}$ from high-latitude data, and \cite{Burke14} estimate a rate of $\sim 2 \times 10^3$~sky$^{-1}$~day$^{-1}$ from intermediate- and low-latitude data. \cite{Burke14} argue that the difference is statistically significant but acknowledge that the estimates use assumptions whose validity about FRBs is unknown. \cite{Rane15} derive a limit of $3.3^{+5.0}_{-2.5} \times 10^3$~day$^{-1}$ sky$^{-1}$ for bursts with a flux density $> 0.1$~Jy at latitudes $|b| < 60\degrees$ and argue that this is consistent with rates from other Parkes surveys. \cite{Karako15}, who did not find any FRBs in GBT350 drift data, derive a limit on the rate of bursts with $S \gtrsim 260$~mJy and widths $\sim 10$~ms at 350~MHz and obtain $\sim 1 \times 10^4$~sky$^{-1}$~day$^{-1}$. 

We assume that FRBs follow Poisson statistics in order to calculate a similar rate limit from AO327-Mock sky coverage. The Poisson probability of detecting exactly $k$ FRBs in a survey of total duration $T$ is 
\be
P(X=k) = \frac{\left(r \theta T\right)^k e^{-\left(r \theta T\right)}}{k!},
\ee
where $\theta$ is the beam area and $r$ is the burst rate. The probability of detecting at least one FRB is $P(X>0) = 1 - P(X=0) = 1 - e^{-\left(r \theta T\right)}$. We calculate that for a 99\% chance of detecting at least one FRB in the 882~h of AO327-Mock data, the rate is $\sim 1 \times 10^5$~sky$^{-1}$~day$^{-1}$ for 10~ms bursts with $S \gtrsim 83$~mJy. The limit derived from AO327-Mock is less stringent than the limits from GBT350 or the Parkes surveys, because AO327-Mock has significantly less total on-sky time and smaller beam size. Since the spectral indices of FRBs are on the whole unknown, we can meaningfully compare the AO327-Mock limit only with the GBT350 limit. AO327-Mock searches 5 times more volume per unit time than GBT350 (\citealt{Deneva13}, Figure~4). Therefore, the AO327-Mock FRB rate limit normalized to the GTB350 search volume is $\sim 2 \times 10^4$~sky$^{-1}$~day$^{-1}$. This estimate will improve when results from AO327-PUPPI are included, in addition to results from reprocessing AO327-Mock data with a DM list tailored for FRB detection. 

\section{Summary}

We have developed Clusterrank, a new algorithm to automatically rank clusters of events recorded by single-pulse searches based on each cluster's likelihood of being generated by a dispersed astrophysical pulse. Clusterrank enabled us to quickly identify 8 RRATs and 14 slow pulsars missed by an FFT-based periodicity search in AO327 drift survey data. The new RRATs have DMs in the range $22.5 - 86.6$~pc~cm$^{-3}$. Five of these sources have period estimates from pulse arrival times; their periods are in the range $0.172 - 3.901$~s. The new pulsars have DMs in the range $23.6 - 133.3$~pc~cm$^{-3}$ and periods in the range $1.249 - 5.012$~s. 

We find that the periods of RRATs found by Clusterrank are drawn from the same distribution as the periods of sources in the RRATalog, and that the periods of pulsars and RRATs discovered by Clusterrank are consistent with having the same underlying distribution. We also find that there is no significant difference between the underlying DM or spatial distributions of new sources found by AO327 via periodicity search vs. new sources found via using Clusterrank on PRESTO single-pulse search output. 

Although we search AO327 data with DMs up to 1000~pc~cm$^{-3}$, we have not yet found any highly dispersed pulses indicative of FRBs. We identify a common optimization in constructing trial DMs lists for pulsar surveys that likely hinders the identification of such pulses either visually or algorithmically and recommend that the DM search space be deliberately oversampled for DM~$\gtrsim 500$~pc~cm$^{-3}$ compared to what is optimal for Galactic sources. 

We thank Ben Arthur and Chen Karako-Argaman for useful discussions. J.S.D. was supported by the NASA Fermi Guest Investigator program and the Chief of Naval Research. M.A.M., M.B., and S.D.B. were supported by NSF award numbers 0968296 and 1327526. The Arecibo Observatory is operated by SRI International under a cooperative agreement with the National Science Foundation (AST-1100968), and in alliance with Ana G. M\'{e}ndez-Universidad Metropolitana, and the Universities Space Research Association.

\end{document}